\documentclass[lettersize,journal]{IEEEtran}

\usepackage[final]{changes} 
\definechangesauthor[name={}, color=blue]{rev}

\usepackage{setspace}  
\usepackage{lineno}    

\IEEEoverridecommandlockouts
\usepackage{cite}
\usepackage{amsmath,amssymb,amsfonts}
\usepackage{algorithmic}
\usepackage{graphicx}
\usepackage{textcomp}
\usepackage{xcolor}
\usepackage{caption}
\usepackage{subcaption}
\usepackage[ruled]{algorithm2e}
\usepackage{caption}
\usepackage{subcaption}
\usepackage{epstopdf}
\usepackage{multirow}
\usepackage{svg}
\usepackage{float}
\usepackage{makecell}

\newcommand{\real}[1]{\text{Re} \left( #1 \right)}
\newcommand{\imag}[1]{\text{Im} \left( #1 \right)}

\newcommand{\sss}{\mathcal{S}}
\newcommand{\Dtr}{\mathcal{D}_{\text{tr}}}
\newcommand{\Dtest}{\mathcal{D}_{\text{test}}}
\newcommand{\Ntr}{N_{\text{tr}}}
\newcommand{\Ntest}{N_{\text{test}}}

\newcommand{\param}{\mathbf{w}}

\def\BibTeX{{\rm B\kern-.05em{\sc i\kern-.025em b}\kern-.08em
    T\kern-.1667em\lower.7ex\hbox{E}\kern-.125emX}}
\begin{document}

\title{Joint Source-Environment Adaptation of Data-Driven Underwater Acoustic Source Ranging Based on Model Uncertainty 
\thanks{This work has been supported by the Office of Naval Research (ONR) under grant N00014-19-1-2662.}}

\author{
\IEEEauthorblockN{Dariush Kari},
\thanks{Manuscript received August 1, 2024; revised September 16, 2024. (Corresponding author: Dariush Kari)}
\thanks{Dariush Kari is with the Interdisciplinary Health Sciences Institute (IHSI), University of Illinois Urbana-Champaign, Urbana, IL 61801, USA (email: dkari2@illinois.edu).}
\and 
\IEEEauthorblockN{Hari Vishnu,~\IEEEmembership{Senior Member,~IEEE,}} \thanks{Hari Vishnu is with Acoustic Research Laboratory, National University
of Singapore, Tropical Marine Science Institute, Singapore 119222 (email: tmshv@nus.edu.sg)}
\and
\IEEEauthorblockN{Andrew C. Singer,~\IEEEmembership{Fellow,~IEEE}} \thanks{Andrew C. Singer is with the Department of Electrical and Computer Engineering, Stony Brook University, Stony Brook, NY 11794, USA (email: andrew.c.singer@stonybrook.edu).}
}

\maketitle

\begin{abstract}
Adapting pre-trained deep learning models to new and unknown environments \replaced[]{remains a major}{is a difficult} challenge in underwater acoustic localization. We show that although \replaced[]{the performance of pre-trained models}{pre-trained models have performance that} suffers from mismatch between the training and test data, they generally exhibit a higher \replaced[]{uncertainty}{``implied uncertainty''} in environments where there is more mismatch. \added[]{Additionally, in the presence of environmental mismatch, spurious peaks can appear in the output of classification-based localization approaches, which inspires us to define and use a method to quantify the ``implied uncertainty'' based on the number of model output peaks.} Leveraging this notion of implied uncertainty, we partition the test samples into \replaced[]{sets with more certain and less certain samples}{more certain and less certain sets}, and implement \replaced[]{a method to adapt the model to new environments by}{an estimation method} using the certain samples to improve the labeling for uncertain samples\replaced[]{}{, which helps to adapt the model}. \replaced[]{Thus, using this efficient method for model uncertainty quantification, we showcase}{We use an efficient method to quantify model prediction uncertainty, and} an innovative approach to adapt a pre-trained model to unseen underwater environments at test time. This eliminates the need for labeled data from the target environment or the original training data. This adaptation is enhanced by integrating an independent estimate based on the received signal energy. We validate the approach extensively using real experimental data, as well as synthetic data consisting of model-generated signals with real ocean noise. The results demonstrate significant improvements in model prediction accuracy, underscoring the potential of the method to enhance underwater acoustic localization in diverse, noisy, and unknown environments.
\end{abstract}



\section{Introduction}

\added[]{\protect\IEEEPARstart{U}{nderwater} acoustic (UWA) source localization refers to determining all or some of the coordinates of an underwater acoustic source using the corresponding recorded signals \cite{baggeroer1988matched,  hursky2001matched, niu2017source, weiss2022semi, kari2023gradient, kari2025mismatch}. Source ranging refers to estimating the distance of a source and can be achieved using conventional methods such as matched field processing (MFP) \cite{baggeroer1988matched, hursky2001matched} or deep learning (DL)-based methods \cite{niu2017source, chen2021model, niu2019deep}}. \replaced[]{}{Deep learning (}DL\replaced[]{}{)}-based \replaced[]{UWA}{underwater acoustic (UWA)} localization algorithms \cite{niu2017source, yangzhou2019deep, chen2021model, wang2018underwater} tend to generalize poorly to mismatched environments\cite{chen2021model, liu2023unsupervised, kari2024joint}. To overcome small mismatches between the testing and training sets, domain adaptation (DA) methods \replaced[]{have been proposed that}{} match some of the statistics of the testing data to those of the training data\cite{ganin2015unsupervised}, which, at inference time, requires access to the training data acquired with known or controlled sources. However, for low-power \replaced[]{or low storage-capacity}{} underwater devices working in a decentralized manner \cite{song2020underwater, zhuo2020auv, zhang2020node}, training data will not always be available at the inference device, either due to the communication cost of transferring large datasets between such devices, or security or privacy issues. Therefore, this paper seeks to improve the generalization performance of DL-based models for UWA localization via \emph{test-time adaptation} (TTA) \footnote{Note that in machine learning parlance, test-time adaptation is also called source-free adaptation, where ``source'' refers to the training domain. However, to avoid confusion with the acoustic source, we use the term test-time adaptation.} \cite{liang2020we}. \par

To provide a more robust solution than matched field processing (MFP) \cite{baggeroer1988matched,sullivan1993estimation}, Chen and Schmidt \cite{chen2021model} propose a model-based convolutional neural network (CNN) that outperforms MFP in the presence of sea depth mismatch. It generates the training data using a propagation model with the sound speed profile (SSP) set according to the values measured in a real experiment. This method ignores the parameter shift between the synthetic training data and the real testing data and assumes accurate prior knowledge of the parameters of the test environment, which is rarely available. This motivates further improvement using an adaptation mechanism.\par

The environmental mismatch problem is known in the machine learning literature as domain shift \cite{zhang2013domain}, and has been tackled with approaches including \replaced[]{DA}{domain adaptation} or generalization \cite{zhang2013domain, liang2020we}, meta learning \cite{zhang2021meta}, transfer learning \cite{wang2019deep, yosinski2014transferable}, and data augmentation \cite{yao2022improving}. Transfer learning requires labels for some of the test data, which are usually not available in UWA problems and encourages pursuit of DA methods. Moreover, although DA and transfer learning have been studied for some cases in underwater acoustic localization \cite{wang2019deep, liu2021deep}, \cite{liu2023unsupervised, long2023deep}, more extensive investigations into DA are warranted.\par

Domain adaptation can leverage  training data whenever available to learn any underlying domain shift. For UWA localization, inspired by deep subdomain adaptation networks\cite{zhu2020deep}, Liu, et al. \cite{liu2023unsupervised} propose an unsupervised DA method that aims to align the feature distributions of training and test data at every layer of the network. Long, et al. \cite{long2023deep} apply an adversarial method based on \cite{ganin2015unsupervised} to extract useful features that are invariant with respect to domain shifts. This approach adds a domain classifier to the original deep model and uses a training loss designed to promote features that confuse this domain classifier in order to achieve domain invariance. These methods, though effective, rely on access to training data. In the absence of training data, TTA is still possible by inferring some useful statistics (features) of the training set from the pre-trained model itself \cite{liang2020we}.\par

There is a rich literature on TTA including source hypothesis transfer (SHOT) \cite{liang2020we}, entropy minimization \cite{wang2021tent}, universal TTA \cite{kundu2020universal}, contrastive TTA\cite{chen2022contrastive}, adaptive adversarial network-based TTA \cite{xia2021adaptive}, and distribution estimation \cite{ding2022source}. To perform TTA, we leverage an implied model uncertainty \cite{gawlikowski2023survey, gal2016dropout, smith2018understanding, abdar2021review}. Although conformal prediction \cite{angelopoulos2021gentle} has recently been adopted for uncertainty quantification in data-driven UWA localization \cite{khurjekar2023uncertainty}, it requires a labeled dataset from the test environment for calibration. To explore the effects of mismatch on model uncertainty without any labeled dataset available at the test time, we adopt model uncertainty using mutual information (MUMI) \cite{smith2018understanding} \replaced[]{as a metric}{}. To avoid the computational burden of MUMI during adaptation, in Section \ref{sec:PU}, we cast the ranging problem as a classification task that can leverage computationally simpler \emph{implied} uncertainty quantification methods such as peakwise uncertainty (PU). Our observations in Section \ref{sec:uncertainty} demonstrate that both MUMI and PU increase with an increase in the amount of mismatch\replaced[]{, validating their utility}{}.\par

Environmental mismatches do not affect all samples to the same extent - some samples are affected more and result in a higher uncertainty. For instance, the signal received by the receiver array from a source at a short distance is dominated by the direct path arrivals and first-order reflected raypaths (single reflection from either surface or bottom) \cite{weiss2022semi}. Thus, the depth mismatch affects farther sources more significantly due to the contribution of higher-order bottom reflections received by the array. We study source ranging problems in which there are diverse test samples from different ranges associated with the same source. By partitioning the test samples into certain and uncertain sets based on their PU scores, we can implicitly infer the source power.\par


Typical data-driven methods in UWA signal processing perform normalization with respect to the received signal strength to obtain source-invariant features \cite{niu2017source}. However, this discounts the information in the received signal strength which can aid UWA localization, as demonstrated by some methods \cite{xu2016rss, zhang2016received}. A pre-trained network deprived of the source power information, as discussed in Section \ref{sec:PU}, will be confused in the presence of environmental mismatches. To reduce the model confusion, we develop another robust and coarse estimation process based on the received signal strength named the joint source-environment adaptation (JSEA). This leverages the set of certain test samples and enhances the generalization across different environments, while retaining the source-invariance property of the pre-trained model.\par


The rest of the paper is organized as follows. Section~\ref{sec:SCM} provides the necessary background on the MFP and CNN approaches for localization. Section~\ref{sec:uncertainty} investigates uncertainty quantification in the UWA localization problem. Section~\ref{sec:adaptation} provides the SHOT approach to adaptation and our proposed method for JSEA. Finally, Section~\ref{sec:simulations} demonstrates the efficacy of the proposed method on several synthetic and real datasets, followed by \added[]{discussions on potential extensions and limitations of JSEA in Section~\ref{sec:ext-lim} and} some concluding remarks in Section~\ref{sec:conclusion}.

\section{Localization Using Sample Covariance Matrices} \label{sec:SCM}
\textbf{Notation:} Bold letters denote matrices and vectors, $H(\mathbf{y})$ indicates the discrete entropy of a probability mass function (PMF) $\mathbf{y}$, $D_{KL}$ denotes the KL-divergence, $\real{x}$ and $\imag{x}$ respectively denote the real and imaginary parts of $x$, \replaced[]{Superscripts $^{\mathsf{T}}$ and $^{\mathsf{H}}$}{$\mathbf{A}^{\mathsf{T}}$ and $\mathbf{A}^{\mathsf{H}}$} denote respectively the transpose and conjugate transpose of the \replaced[]{matrices or vectors}{matrix or vector $\mathbf{A}$}. For a set of test points $\sss$, the complement is \replaced[]{represented}{shown} by $\sss^\mathsf{c} = \Dtest - \sss$ and the cardinality is shown by $|\sss|$.\par

Sample covariance matrices (SCM) are usually used for UWA array-based localization because they are sufficient statistics for the jointly Gaussian signal and noise \cite{gerstoft2020parametric} scenarios. The normalized SCM of an $L$-element array at a frequency $f$ using $P$ snapshots $\mathbf{\Tilde{r}}^{(p)}(f)$ is calculated as 
\begin{equation}\label{eq:SCM}
\tilde{\mathbf{C}}(f) = \frac 1 P \sum_{p=1}^{P} \mathbf{\Tilde{r}}^{(p)}(f) \mathbf{\Tilde{r}}^{(p)}(f)^{\mathsf{H}},
\end{equation}
where $\mathbf{\Tilde{r}}^{(p)}(f) = [ \Tilde{r}^{(p)}_1(f),  \Tilde{r}^{(p)}_2(f), ...,  \Tilde{r}^{(p)}_L(f)]^\mathsf{T}$ and
\begin{align}
    \Tilde{r}^{(p)}_l(f) &= \frac{r^{(p)}_l(f)}{\sqrt{\sum_{l=1}^{L} |r_l^{(p)}(f)|^2}}, \quad l \in \{1,2,...,L \},
\end{align}
where the scalar $r^{(p)}_l(f)$ is the complex Fourier coefficient of the $p$-th segment of the received signal on the $l$-th hydrophone at the frequency $f$. For the synthetic data in the simulations, we use the \replaced[]{signals generated by the KRAKEN normal-mode model\cite{porter1992kraken}}{KRAKEN-generated values} for $\Tilde{r}^{(p)}_l(f)$, while for the real data, we take the Fourier transform of the recorded temporal waveform \replaced[]{and use the Fourier coefficients at the frequency of interest}{}, as explained in Section \ref{sec:SWellEx}.\par

\subsection{MFP approach}
Given that the acoustic source \replaced[]{emits}{propagates} a narrowband signal at a frequency $f$, we use the Bartlett processor \cite{gemba2017adaptive} as an MFP baseline, which is defined as
\begin{equation}\label{eq:MFP}
    \hat{d} = \arg \max_{d} \mathbf{\Tilde{r}}(f,d)^{\mathsf{H}} \Tilde{\mathbf{C}}(f) \mathbf{\Tilde{r}}(f,d),
\end{equation}
where $\Tilde{\mathbf{C}}$ is the normalized SCM of the measured data, \replaced[]{and}{} $\mathbf{\Tilde{r}}(f,d)$ is the replica field at frequency $f$ generated by a source at range $d$. For each $d$, the corresponding $\mathbf{\Tilde{r}}(f,d)$ (of the CNN approach) from the training set is used at  inference time.
\subsection{CNN approach}
Localization can be formulated as a regression or a classification problem \cite{chen2021model,weiss2022direct}. Classification models usually assume equidistant labels, as opposed to the regression models that preserve the distance among labels. To address this issue, rather than predicting the true class, we train the classification network to predict a softened label, viz. a PMF, based on the true class. This metric-inspired label softening preserves distances between different classes, makes the model more amenable to the localization problem, and facilitates the proposed uncertainty-based TTA approach. The proposed classification and regression models are both based on a CNN architecture, differing only in the last layer, as in \cite{chen2021model}.\par

\begin{figure}[ht]
\centering
    \includegraphics[width= \linewidth]{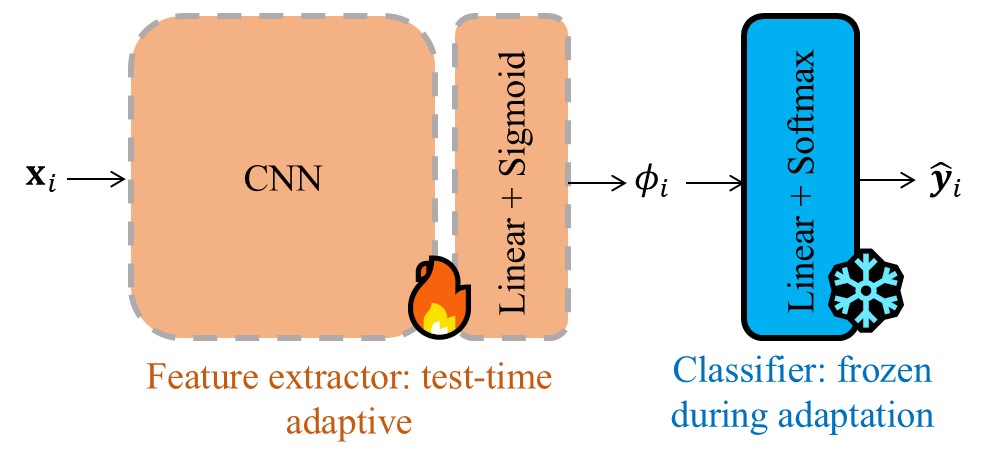}
    \caption{The CNN classifier includes a feature extraction part followed by a linear classifier that will be frozen during adaptation.}
    \label{fig:sfda}
\end{figure}

\begin{figure}[htb]
\centering
    \includegraphics[width= \linewidth]{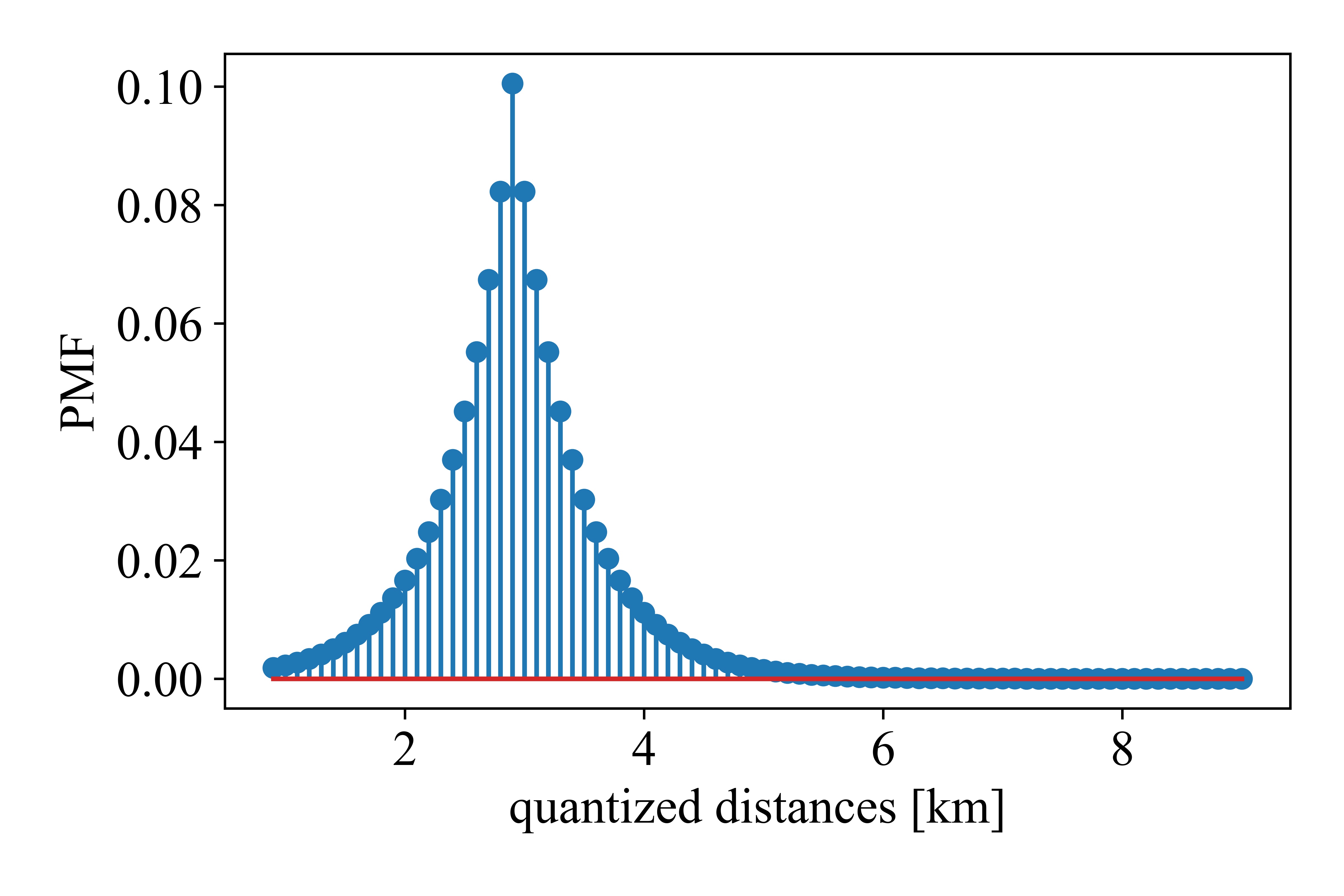}
    \caption{An example of a metric-inspired smoothed label $\mathbf{y}_i$ corresponding to $d_i = 2.9$~km or $d^q_i = 20$. Here, we have used the absolute error as the metric, hence, we have used a truncated exponential as the label. The quantization bins are $100$ m wide.}
    \label{fig:smoothed}
\end{figure}

The classifier CNN architecture is detailed as follows. As depicted in Fig. \ref{fig:sfda}, the feature extractor part of the model consists of a CNN that takes the real and imaginary parts of the SCM, $\mathbf{x}_i = \left[\real{\Tilde{\mathbf{C}_i}}, \imag{\Tilde{\mathbf{C}_i}}\right]$, as a $2$-channel input (a real tensor of dimension $2 \times L \times L$) and a linear layer that generates the feature vector $\boldsymbol{\phi}_i$. These features are then fed into a linear classifier to generate the output PMF $\hat{\mathbf{y}}_i$. Denoting the region of interest for the source range by $[D_{\min}, \; D_{\max}]$, for each training sample $i$, we quantize the range $d_i$ by $d^q_i = \lfloor \frac{d_i - D_{\min}}{B} + 0.5 \rfloor$ to obtain classification outputs with a resolution of  $B=100$~m (and consequently, a PMF with bin size $B=100$~m). A one-hot representation of $d^q_i$ is used in classification tasks. However, since we use mean absolute error (MAE) to evaluate our results, we define the soft label $ \mathbf{y}_i = [y_{i1}, y_{i2}, ..., y_{iM}] = \eta(d^q_i)$ as
\begin{align}
    y_{ik} = \frac {\exp (-|k - d^q_i|/\sigma)}{\sum_{k=1}^M \exp (-|k - d^q_i|/\sigma)},
\end{align}
where $M$ is the number of classes and $\sigma$ is a hyperparameter determining the spread of the peak in the soft label. A small $\sigma$ effectively renders the labels one-hot coded while a large $\sigma$ decreases the accuracy by encouraging high-entropy output labels. We empirically conclude that $1 \leq \sigma \leq 10$ is appropriate for quantization bins of size $B = 100$ m. Fig. \ref{fig:smoothed} shows an example of a soft label with $\sigma = 5$.

Therefore, the training dataset is denoted by $\Dtr = \{(\mathbf{x}_i, \mathbf{y}_i)\}_{i=1}^{\Ntr}$ and the test set that does not include labels is denoted by $\Dtest = \{\mathbf{x}_i\}_{i=1}^{\Ntest}$. We use the loss function $\mathcal{L}_{tr} = \frac 1 {\Ntr} \sum_{i=1}^{\Ntr} \mathsf{CE}(\mathbf{y}_i, \hat{\mathbf{y}}_i)$ for training, where $\hat{\mathbf{y}}_i$ is the network output and $\mathsf{CE}$ is the cross-entropy loss defined by
\begin{equation}
    \mathsf{CE}(\mathbf{y}_i, \hat{\mathbf{y}}_i) \triangleq - \sum_{k=1}^M y_{ik} \log \hat{y}_{ik}.
\end{equation}

The CNN structure used herein is similar to that presented in \cite{chen2021model} with a $2$-channel input and $3$ convolutional layers with $6$, $38$, and $40$ channels and kernels of sizes $3$, $5$, and $5$, respectively. The output of the feature extraction is a $256$-dimensional vector $\boldsymbol{\phi}_i$, which is then fed into the classifier to generate $\hat{\mathbf{y}}_i$.

\section{Model Uncertainty Under Mismatch} \label{sec:uncertainty}
For TTA, we first compute the uncertainty of the DL model about its estimations, and then implicitly extract the power of the acoustic source that generated the test data, from those test samples about which the model is most certain. We will discuss two uncertainty quantification methods that can be used with classification or regression models. Although localization naturally fits into a regression paradigm \cite{weiss2022direct}, regression models in their standard form only output point estimates. A stochastic regression model (e.g., a dropout network) can produce a PMF output by several realizations of the forward pass \cite{smith2018understanding}, which can then be used in MUMI or PU. However, the deterministic classification model already outputs a PMF in its forward pass, which can be used in the PU method.

\subsection{Model Uncertainty using Mutual Information (MUMI)}\label{sec:mumi}
There are two different sources of uncertainties in data-driven models \cite{smith2018understanding, abdar2021review}: 1) epistemic that corresponds to the lack of training data and 2) aleatoric that corresponds to noisy or inherently ambiguous data. To alleviate the effects of mismatch between the training and test data, we focus on the epistemic uncertainty. Consider a dropout network with the weights vector $\param$. Treating $\param$ as a random variable, the network output for a given test sample $\mathbf{x}_{\mathsf{test}} = \left[\real{\Tilde{\mathbf{C}}_{\mathsf{test}}}, \imag{\Tilde{\mathbf{C}}_{\mathsf{test}}}\right]$ is also a random variable. If each realization of $\param$ yields a different output for the same input $\mathbf{x}_{\mathsf{test}}$, the model is considered uncertain about the output. MUMI \cite{smith2018understanding} quantifies the amount of information that we would obtain had we received the true label (acoustic source range) $d_{\mathsf{test}}$ for the corresponding $\mathbf{x}_{\mathsf{test}}$. This is represented in terms of the mutual information \cite{cover1999elements} $I(d_{\mathsf{test}};\param | \Dtr, \mathbf{x}_{\mathsf{test}})$, where the conditioning over $\Dtr$ is because we have used the training data to obtain $\param$. Expanding this mutual information yields
\begin{align} \label{eq:MUMI_original}
    I(d_{\mathsf{test}};\param | \Dtr, \mathbf{x}_{\mathsf{test}}) & = H\left(p(d_{\mathsf{test}}|\Dtr, \mathbf{x}_{\mathsf{test}})\right) \nonumber \\
    & - \mathbb{E}_{p(\param | \Dtr)} \left[ H \left( p(d_{\mathsf{test}}|\Dtr, \param, \mathbf{x}_{\mathsf{test}}) \right) \right],
\end{align}
where for a given function $h$, $\mathbb{E}_{p(\param | \Dtr)}[h(\param)]$ denotes the expectation of the random variable $h(\param)$ when $\param$ is distributed according to the probability density function $p(\param | \Dtr)$. Although computing $\mathbb{E}_{p(\param | \Dtr)}[h(\param)]$ is intractable in general, a dropout approximating distribution $q_{\theta}(\param | \Dtr)$ can efficiently approximate $p(\param | \Dtr)$ \cite{smith2018understanding}, where $\theta$ represents the parameters of a dropout neural network and each realization of the dropout mask corresponds to a sampling from $q_{\theta}(\param|\Dtr)$ that yields a different weight vector $\param_j$. Accordingly, the label likelihood $p(d_{\mathsf{test}}|\Dtr, \mathbf{x}_{\mathsf{test}})$ in \eqref{eq:MUMI_original} can be computed by marginalization over $\param$ and a Monte-Carlo approximation as
\begin{align}\label{eq:H_of_E}
p(d_{\mathsf{test}}|\Dtr, \mathbf{x}_{\mathsf{test}}) & \approx \int p(d_{\mathsf{test}}| \param, \Dtr, \mathbf{x}_{\mathsf{test}}) q_{\theta}(\param|\Dtr) d \param \nonumber \\ 
    & \triangleq \frac 1 J \sum_{j=1}^J \hat p(\hat{y}_{j}|\param_j, \Dtr, \mathbf{x}_{\mathsf{test}}) \nonumber \\
    & = \frac 1 J \sum_{j=1}^J \hat p(\hat{y}_{j}|\param_j, \mathbf{x}_{\mathsf{test}}),
\end{align}
where $\param_j \sim q_{\theta}(\param|\Dtr), j=1, ..., J$ are weights sampled from the dropout network with $J$ different dropout realizations, $\hat{y}_{j}$ is the network output corresponding to the weight vector $\param_j$, and $\hat p$ is the empirical likelihood of the model output. Since the training data has been used to obtain $\param$, there is a Markov chain $\Dtr - \param - \hat y_j$, hence, $\hat p(\hat y_j|\Dtr, \param_j, \mathbf{x}_{\mathsf{test}}) = \hat p(\hat y_j|\param_j, \mathbf{x}_{\mathsf{test}})$. Similarly, $p(d_{\mathsf{test}}| \param, \Dtr, \mathbf{x}_{\mathsf{test}})$ can be approximated with $\hat p(\hat y_j|\param_j, \mathbf{x}_{\mathsf{test}})$, which results in

\begin{align} \label{eq:E_of_H}
    \mathbb{E}_{p(\param | \Dtr)} & \left[ H \left( p(d_{\mathsf{test}}|\Dtr, \param, \mathbf{x}_{\mathsf{test}}) \right) \right] \nonumber \\
    & \approx \frac 1 J \sum_{j=1}^J H\left( \hat p(\hat y_j|\param_j, \mathbf{x}_{\mathsf{test}}) \right).
\end{align}
Substituting \eqref{eq:H_of_E} and \eqref{eq:E_of_H} in \eqref{eq:MUMI_original} yields
\begin{align*}
    I(d_{\mathsf{test}};\param | \Dtr, \mathbf{x}_{\mathsf{test}}) & \approx H\left(\frac 1 J \sum_{j=1}^J \hat p(\hat{y}_{j}|\param_j, \mathbf{x}_{\mathsf{test}}) \right) \\ & \quad - \frac 1 J \sum_{j=1}^J H(\hat p(\hat{y}_{j}|\param_j, \mathbf{x}_{\mathsf{test}})).
\end{align*}

Note that MUMI can be used for both classification and regression networks. While in a classification network the output vector $\hat{\mathbf{y}}_j$ is effectively $\hat p(\hat{y}_{j}|\param_j, \mathbf{x}_{\mathsf{test}})$, to obtain the empirical likelihood $\hat p(\hat{y}_{j}|\param_j, \mathbf{x}_{\mathsf{test}})$ in a regression model, we quantize and one-hot encode the output of the regression network, leading to $H(\hat p(\hat{y}_{j}|\param_j, \mathbf{x}_{\mathsf{test}})) = 0$. Thus, the MUMI reduces to 
\begin{equation} \label{eq:MUMI_final}
I(d_{\mathsf{test}};\param | \Dtr, \mathbf{x}_{\mathsf{test}}) \approx H\left(\frac 1 J \sum_{j=1}^J \hat p(\hat{y}_{j}|\param_j, \mathbf{x}_{\mathsf{test}}) \right).
\end{equation}

\begin{figure*}[t]
    \centering
    \begin{subfigure}[b]{0.45\textwidth}
        \centering
        \includegraphics[width=\textwidth]{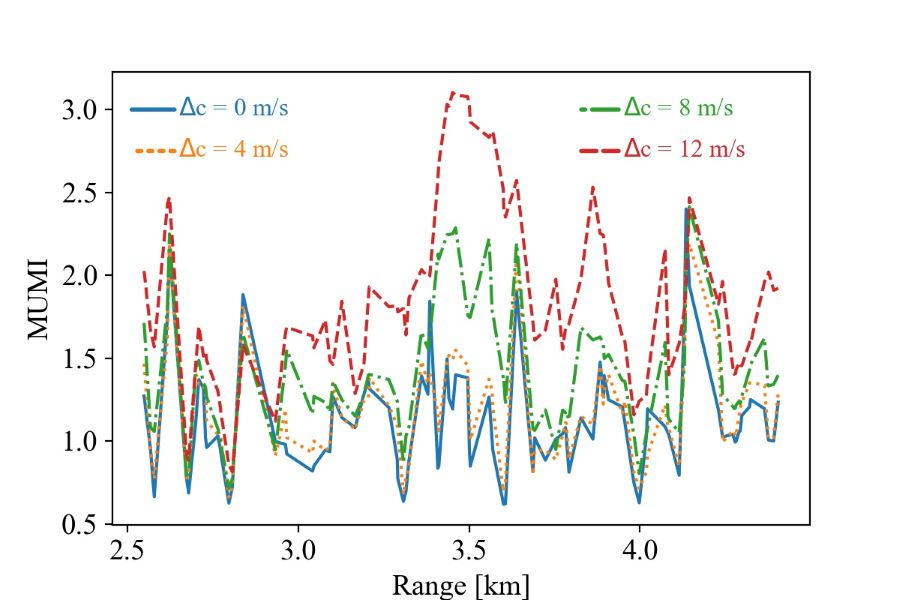}
        \caption{Uncertainty under SSP mismatch.}
        \label{fig:mumi_ssp_example}
    \end{subfigure}
    \hfill
    \begin{subfigure}[b]{0.45\textwidth}
        \centering
        \includegraphics[width=\textwidth]{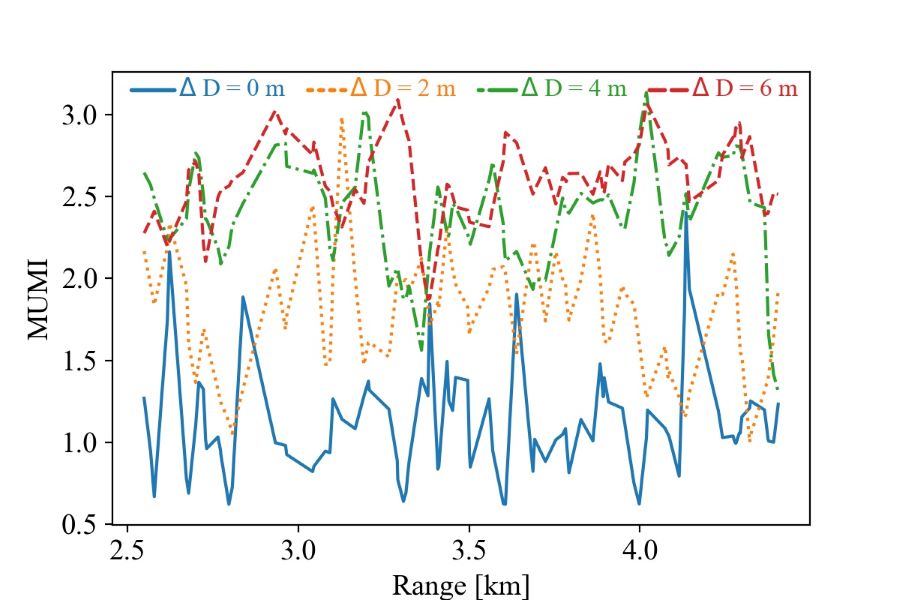}
        \caption{Uncertainty under depth mismatch.}
        \label{fig:mumi_depth_example}
    \end{subfigure}
    \vskip\baselineskip
    \begin{subfigure}[b]{0.45\textwidth}
        \centering
        \includegraphics[width=\textwidth]{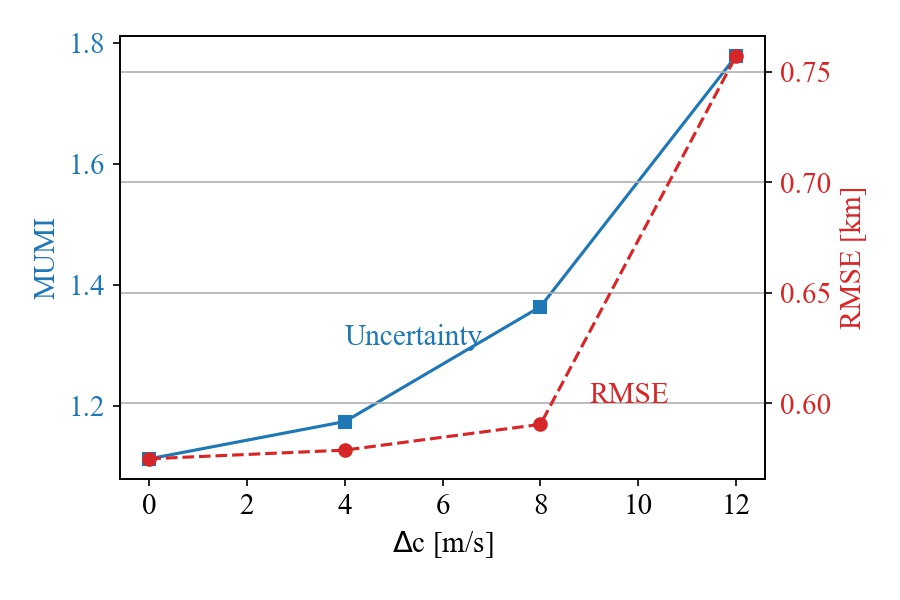}
        \caption{Average uncertainty \added{over all ranges considered} under SSP mismatch.}
        \label{fig:mumi_rmse_ssp}
    \end{subfigure}
    \hfill
    \begin{subfigure}[b]{0.45\textwidth}
        \centering
        \includegraphics[width=\textwidth]{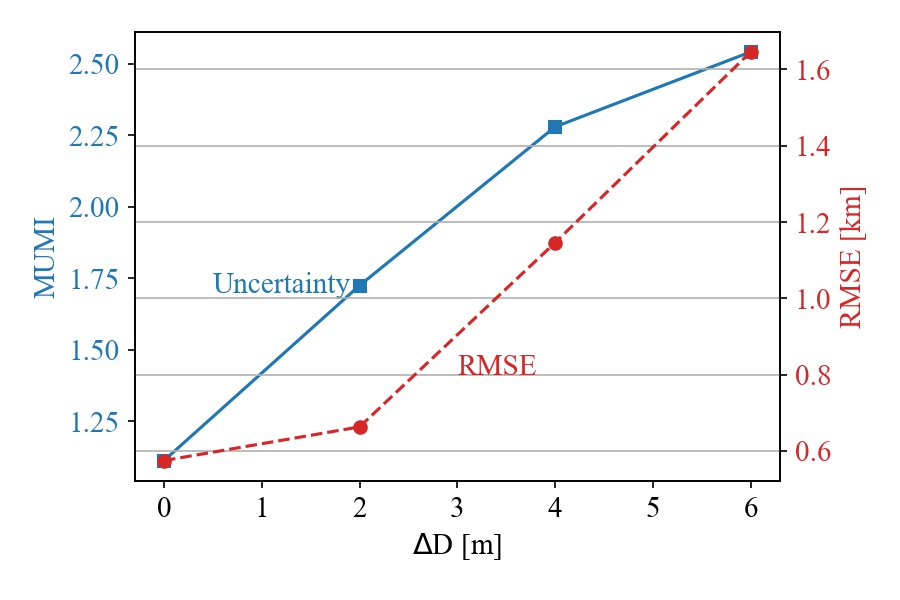}
        \caption{Average uncertainty \added{over all ranges considered} under depth mismatch.}
        \label{fig:mumi_rmse_depth}
    \end{subfigure}
    \caption{MUMI uncertainty of a regression dropout network reveals the mismatch between the training and test environments. These are using the SWellEx-96-inspired synthetic data, consisting of a shallow source with a monotone signal at $f=109$ Hz and all of the results are averaged over $100$ independent realizations of an additive white Gaussian noise (AWGN) channel. Top row shows that the uncertainty is different for each \replaced[]{source range}{sample} and the bottom row shows that, on average, both the uncertainty and the root-mean-squared-error (RMSE) increase with the amount of mismatch.}
    \label{fig:MUMI}
\end{figure*}

Fig. \ref{fig:MUMI} shows that when presented with samples from a mismatched environment, a pre-trained  regression model will have a higher MUMI. This shows that the average MUMI increases with the increase in depth mismatch or SSP mismatch. This behavior is similar to the increased MUMI of DL models during adversarial attacks discussed in \cite{smith2018understanding}. Also, MUMI is seen to be a monotonically increasing function of the \replaced[]{amount of mismatch, validating its use as a metric of uncertainty}{mismatch amount}.

\subsection{Peakwise Uncertainty (PU)} \label{sec:PU}
\begin{figure}[bt]
    \centering
    \begin{subfigure}[b]{0.45\textwidth}
        \centering
        \includegraphics[width=\textwidth]{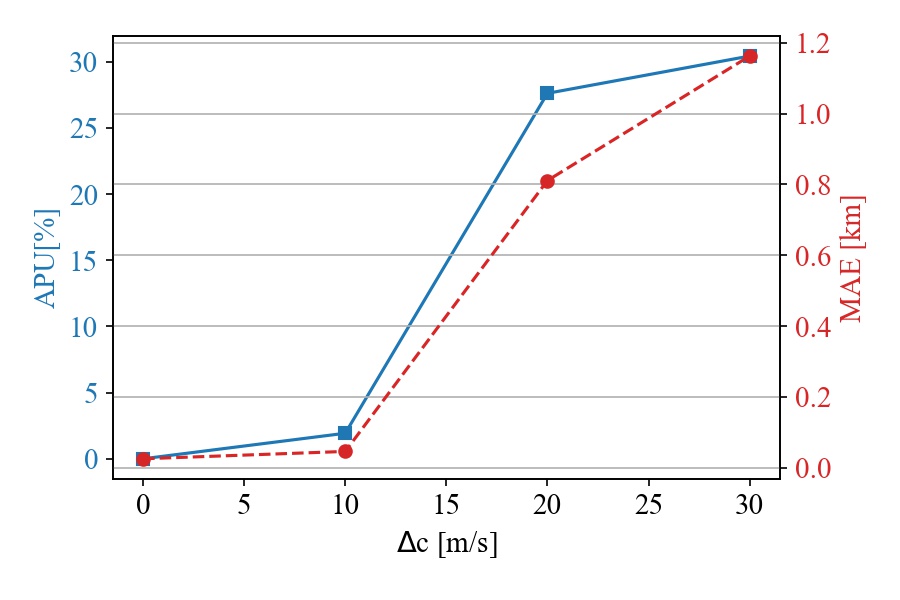}
        \caption{Uncertainty under SSP mismatch.}
        \label{fig:pu_mae_ssp}
    \end{subfigure}
    \hfill
    \begin{subfigure}[b]{0.45\textwidth}
        \centering
        \includegraphics[width=\textwidth]{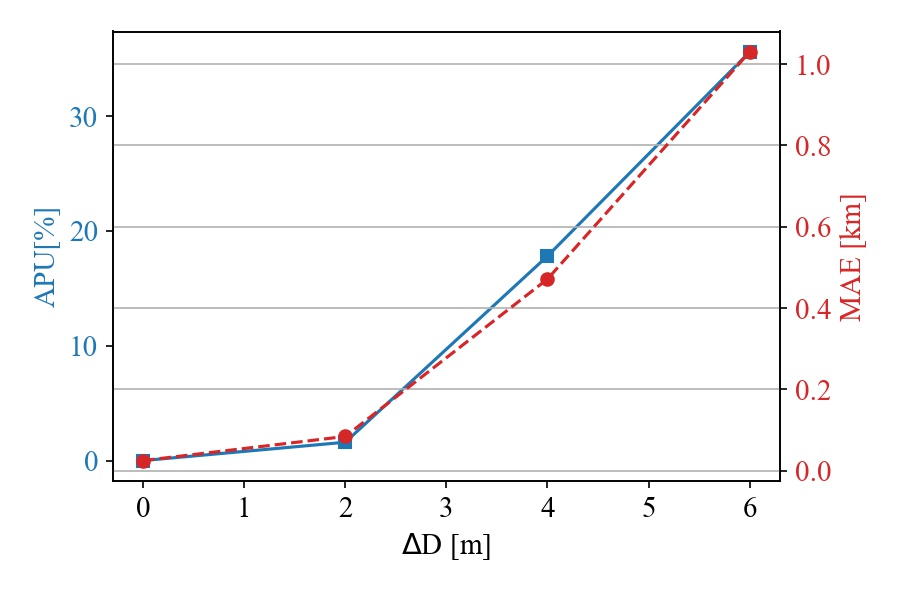}
        \caption{Uncertainty under depth mismatch.}
        \label{fig:pu_mae_depth}
    \end{subfigure}
    \caption{Variation of the \added{average} peakwise uncertainty with the amount of environmental mismatch in (a) depth and (b) SSP.}
    \label{fig:PU}
\end{figure}

While MUMI is theoretically motivated, the MC-dropout requires $J$ realizations of the dropout layers to obtain $\{\hat p(\hat{y}_{j}|\param_j, \mathbf{x}_{\mathsf{test}})\}_{j=1}^J$ which is not feasible for underwater low-power applications. Here, we propose a simpler method to assess uncertainty in the classification approach tailored to the localization problem. This method counts the number of significant peaks in the output PMF, and \replaced[]{is based on the intuition}{asserts} that the model is certain about the corresponding result if there is only one significant peak\replaced[]{, whereas in cases where noise or environmental mismatch dominate, there would not be one significant peak only.}{}Following this approach, we define a ``significant peak" as a peak that is either the only \replaced[]{one}{peak} or is larger than the next-largest peak by a factor \replaced[]{of at least }{}$Q$. The hyperparameter $Q$ should be selected to draw a trade-off between the desired size of $\sss$ and the minimum confidence desired in pseudo labels selected. Thus
\begin{equation}\label{eq:PU}
    PU(\mathbf{x}) = \begin{cases}
        0 \quad \text{if } \hat{\mathbf{y}} \text{ has only 1 significant peak} \\
        1 \quad \text{if } \hat{\mathbf{y}} \text{ has more than 1 significant peak}
    \end{cases}
\end{equation}
where $\hat{\mathbf{y}}$ is the classification network output probability vector. Throughout the paper, we use $Q = 10$. The average PU indicates the percentage of samples about which the model is uncertain \added[]{and is defined as 
    \[APU[\%] = 100 \frac{\sum_{i=1}^N PU(\mathbf{x}_i)}{N}.\]}
    Fig. \ref{fig:PU} shows that the \replaced[]{APU}{average PU} increases as the amount of mismatch increases.

\begin{figure}[htb]
\centering
    \includegraphics[width= \linewidth]{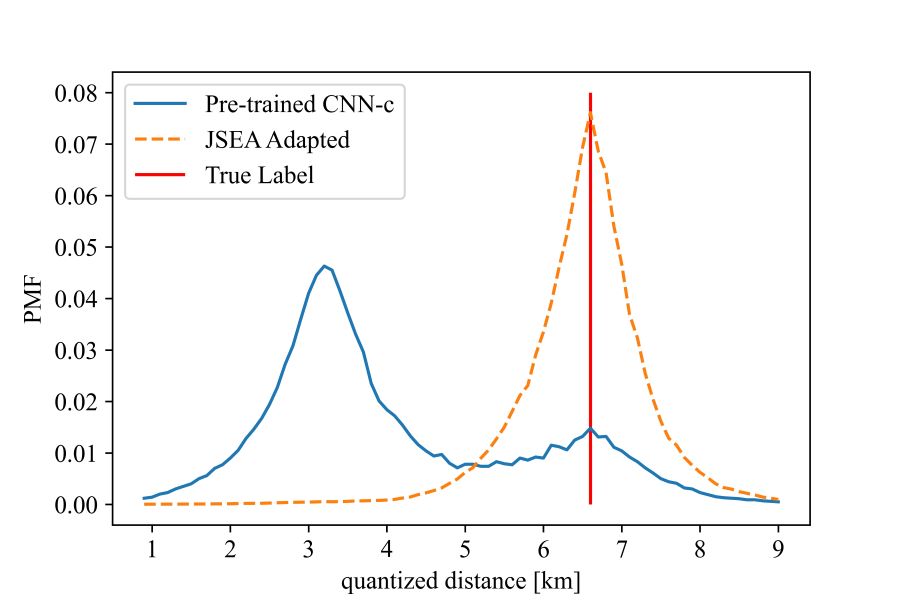}
    \caption{An example of a predicted label in a test environment with a different SSP than the training environment. The quantization bins are $100$ m wide. Observe that the pre-trained model prediction (the label with maximum probability) is far from the ground-truth. However, there is another peak in the output closer to the ground-truth. \replaced[]{After JSEA-based adaptation, the output shows one clear peak at the true label, showing how JSEA corrects and adapts the network to give correct labels in the presence of environmental mismatch}{}.}
    \label{fig:label_example}
\end{figure}

\added[]{Peak counting is not commonly used for uncertainty quantification. Nevertheless, APU shows a similar variation as MUMI in the presence of mismatches. This motivates interpreting PU as an \emph{implied} uncertainty.} As illustrated in the example in Fig. \ref{fig:label_example}, the largest peak in the output of the pre-trained model is not necessarily aligned with the true label. The JSEA algorithm introduced in the next section, determines which peak to select.\par

\section{Test Time Adaptation}\label{sec:adaptation}
Our proposed TTA method, similar to SHOT \cite{liang2020we}, assumes that the DL-based model consists of a domain-dependent feature extractor and a domain-independent classifier (hypothesis). During inference, while freezing the classifier, SHOT fine-tunes the feature extractor to produce confident outputs and prevents it from a degenerate solution using an information maximization loss \cite{liang2020we} that promotes diverse output labels. Although this procedure works well for one-hot coded labels, it must be adjusted for ranging tasks, where labels are not equidistant.

\subsection{Environmental Adaptation Using SHOT}
When presented with data from an environment that is markedly different from the one on which it was trained, the network makes \emph{uncertain} predictions, i.e., it is likely to be more confused between different classes, which leads to inaccurate predictions. During inference, the absence of test labels makes fully-supervised adaptation impossible. Nevertheless, encouraging the model to more confidently preserve its predictions on samples that match to the training data, helps it overcome slight mismatches. To this end, we select a self-supervising subset $\sss$ of the samples and use the pre-trained network's output on these samples as their pseudo-labels, for which we use a loss term $\mathsf{CE}(\mathbf{y}_i^{\mathsf{pseudo}}, \hat{\mathbf{y}}_i)$. Here, $\mathbf{y}_i^{\mathsf{pseudo}}$ is the softened label based on the output of the pre-trained classifier. However, this loss and approach can create a degenerate situation where the network only predicts classes that exist in $\sss$ \cite{liang2020we}. Adding a regularizing loss term to maximize the entropy of the outputs average $\bar{\mathbf{y}} = \frac 1 {\Ntest} \sum_{i=1}^{\Ntest} \hat{\mathbf{y}}_i$ can prevent this situation.\par

At test time, we freeze the pre-trained classifier as depicted in Fig. \ref{fig:sfda}, which predicts the label $\hat{\mathbf{y}}_i = [\hat y_{i1}, \hat y_{i2}, ..., \hat y_{iM}]$ by applying the Softmax() operation to $\boldsymbol{\omega}_k^\mathsf{T} \boldsymbol{\phi}_i$, \replaced[]{defined as}{i.e.},
\begin{equation}
\hat y_{ik} = \frac{\exp(\boldsymbol{\omega}_k^\mathsf{T} \boldsymbol{\phi}_i)}{\sum_{k=1}^M \exp(\boldsymbol{\omega}_k^\mathsf{T} \boldsymbol{\phi}_i)},
\end{equation}
\replaced[]{where}{and} $\boldsymbol{\omega}_k$ is the classification weight vector corresponding to the $k$-th class. Note that if the feature vector $\boldsymbol{\phi}_i$ is highly aligned with $\boldsymbol{\omega}_{k_0}$ for only one $k_0$, then the network predicts a unimodal soft label $\hat{\mathbf{y}}_i$, as opposed to situations where the network generates a multimodal output and is not certain about the predicted label. Inspired by this observation, we construct the self-supervising subset $\sss$ according to
\begin{equation} \label{eq:SSS}
\sss = \{i \; : 1 \leq i \leq \Ntest \, \, \text{s.t.} \, \, PU(\mathbf{x}_i)=0 \}.
\end{equation}
\par

\begin{figure*}
        \centering
        \includegraphics[width=\linewidth]{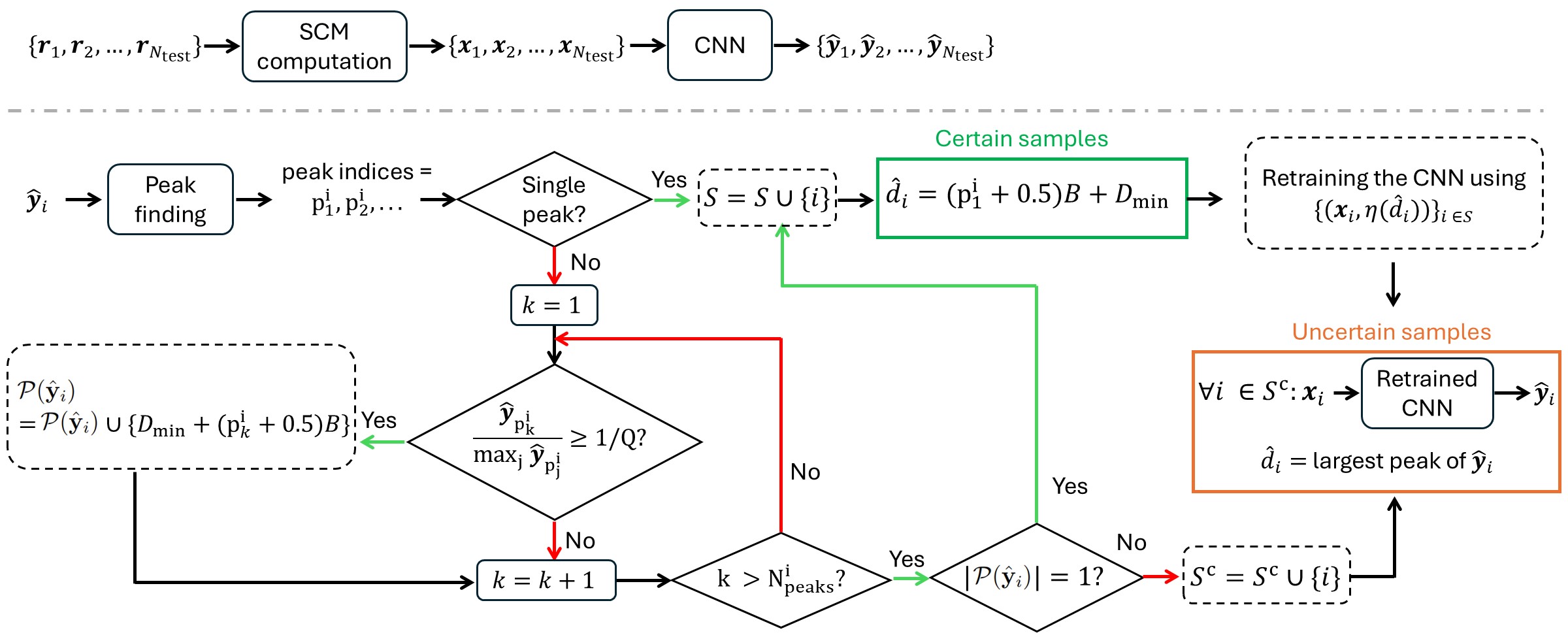}
        \caption{\added[]{A flowchart for the SHOT approach for adapting the localization model. There are three major steps in SHOT: First, the CNN model estimates the labels; then the set of certain samples $\sss$ is determined. For the samples in $\sss$, the CNN estimates are considered the final range estimates (pseudo-labels). Then, the CNN is retrained using the pseudo-labeled samples in $\sss$. Finally, for each of the uncertain samples in $\sss^c$, the final range estimate is obtained from the retrained CNN.}}
        \label{fig:SHOT-flowchart}
\end{figure*}

To perform the adaptation, we minimize the SHOT loss defined by
\begin{align}\label{eq:SHOT_loss}
    \mathcal{L}_{\text{SHOT}} &= - H \left( \bar{\mathbf{y}} \right) + \frac \beta {|\sss|} \sum_{i \in \sss} \mathsf{CE}(\mathbf{y}_i^{\mathsf{pseudo}}, \hat{\mathbf{y}}_i),
\end{align}
where $\beta$ is a positive hyperparameter. We then employ the Adam optimizer \cite{Kingma2015Adam} to minimize $\mathcal{L}_{\text{SHOT}}$ with a step size of $\mu_\text{SHOT}$\replaced[]{ to update the CNN feature extractor}{}. The cross-entropy term encourages the network to remain certain about its existing confident predictions, while increasing $H \left( \bar{\mathbf{y}} \right)$ prevents it from assigning all inputs to the same output by encouraging diverse outputs \cite{liang2020we}. \added[]{Fig. \ref{fig:SHOT-flowchart} provides a flowchart for adapting the model using the SHOT algorithm.}

Although effective at rectifying errors due to small mismatches, SHOT does not directly enhance the estimates for samples in $\sss^c$ and does not focus on removing the model's confusions. To alleviate this, we now introduce another adaptation mechanism that explicitly selects pseudo-labels for the samples in $\sss^c$.

\subsection{Joint Source-Environment Adaptation (JSEA)}

\begin{figure}
    \centering
    \includegraphics[width=\linewidth]{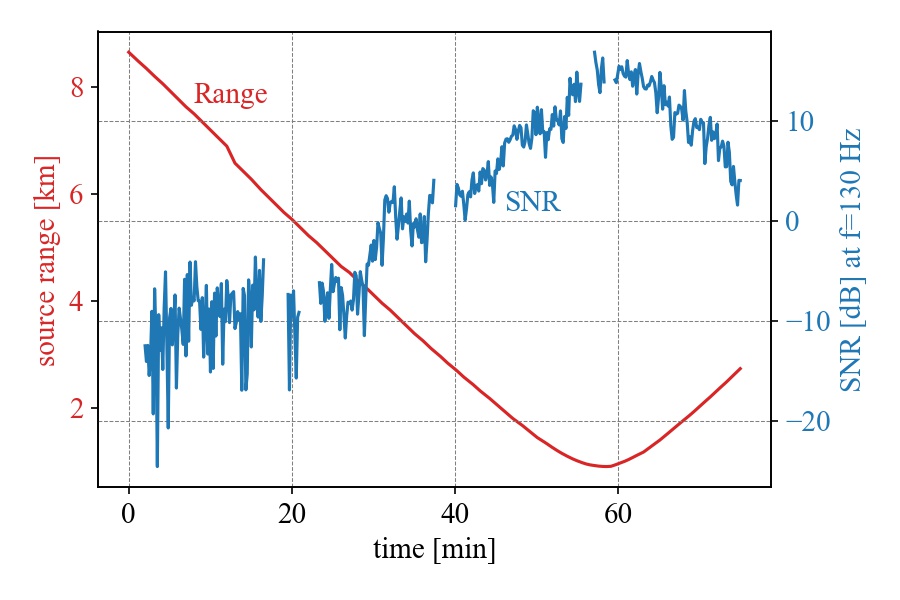}
    \caption{The received SNR (blue) on the vertical linear array at frequency $f=130$ Hz, emitted from the deep source in the event S5 of the SWellEx-96 experiment \cite{Booth2015SWellex}, and source range from the receiver (red). The gaps in the SNR plot indicate some time intervals during which transmission was stopped and the array recorded only the ambient noise. These were used for the SNR calculation.}
    \label{fig:SWellEx-power}
\end{figure}

While TTA methods in the machine learning literature focus on \emph{domain} adaptation, they do not consider or distinguish the \emph{source} that generates data and the \emph{environment} where data is generated. In UWA, however, there is a clear distinction between these two factors. To improve the generalization performance of data-driven methods in UWA signal processing, a necessary step is normalization with respect to the source, which effectively yields source\replaced[]{ power}-invariant features and better generalization across different sources. Accordingly, array-based localization usually involves normalization by the total received signal power at the array elements. \replaced[]{However, by doing this, one loses valuable information on the source power that could be leveraged to improve localization. For example, as shown in}{According to} Fig. \ref{fig:SWellEx-power}, the array SNR at the given frequency can provide a range estimate that (i) is coarse, due to the local fluctuations in the signal power level, but (ii) is robust because the general trend is that higher SNR is associated with a closer source, regardless of the environment. To exploit the distinction between the source and the environment, we implicitly estimate the test source power from a limited set of reliable samples (samples with a unimodal output PMF). We then develop a signal strength-based approach to estimate the range of more difficult samples (samples with a multimodal output PMF). This results in combining the DL-based and signal strength-based approaches in a Bayesian manner to fully use the information in an efficient manner. \par

The method is detailed as follows. Given that $\mathbf{C} = \psi \; \tilde{\mathbf{C}}$, where $\psi = \sum_{l=1}^{L} |r_l^{(p)}(f)|^2$ (which is an accurate definition of received signal power for $P=1$ snapshot case and an approximate estimate for $P>1$), and assuming that $\psi$ and $\tilde{\mathbf{C}}$ are independent, by Bayes' theorem
\begin{align}
    p(d | \mathbf{C}) &= p(d | \tilde{\mathbf{C}}, \psi) = \frac{p(\tilde{\mathbf{C}}, \psi | d) p(d)}{p(\tilde{\mathbf{C}}) p(\psi)} \nonumber \\
    &= \frac{p(\tilde{\mathbf{C}} | \psi, d) p(\psi|d) p(d)}{p(\tilde{\mathbf{C}}) p(\psi)} = \frac{ p(d | \tilde{\mathbf{C}}) p(d | \psi)}{p(d)}.
\end{align}
As a result, if $d$ is uniformly distributed over the parameter space of interest, we have $p(d | \mathbf{C}) \propto p(d | \tilde{\mathbf{C}}) p(d | \psi)$. The received signal power $\psi$ depends on the source power that can be estimated from $\sss$ in \eqref{eq:SSS}. Note that both $p(d | \tilde{\mathbf{C}})$ and $p(d | \psi)$ can be used for localization: $p(d | \tilde{\mathbf{C}})$ provides a finer estimate based on the normalized SCM (and typically more sensitive to mismatches), while $p(d | \psi)$ provides a coarser estimate based on the received signal power that is more robust to environmental mismatches. Since both estimates use different and independent pieces of information, by exploiting both terms we can significantly improve the performance of localization algorithms in the presence of mismatches. \par

\begin{figure*}
        \centering
        \includegraphics[width=\linewidth]{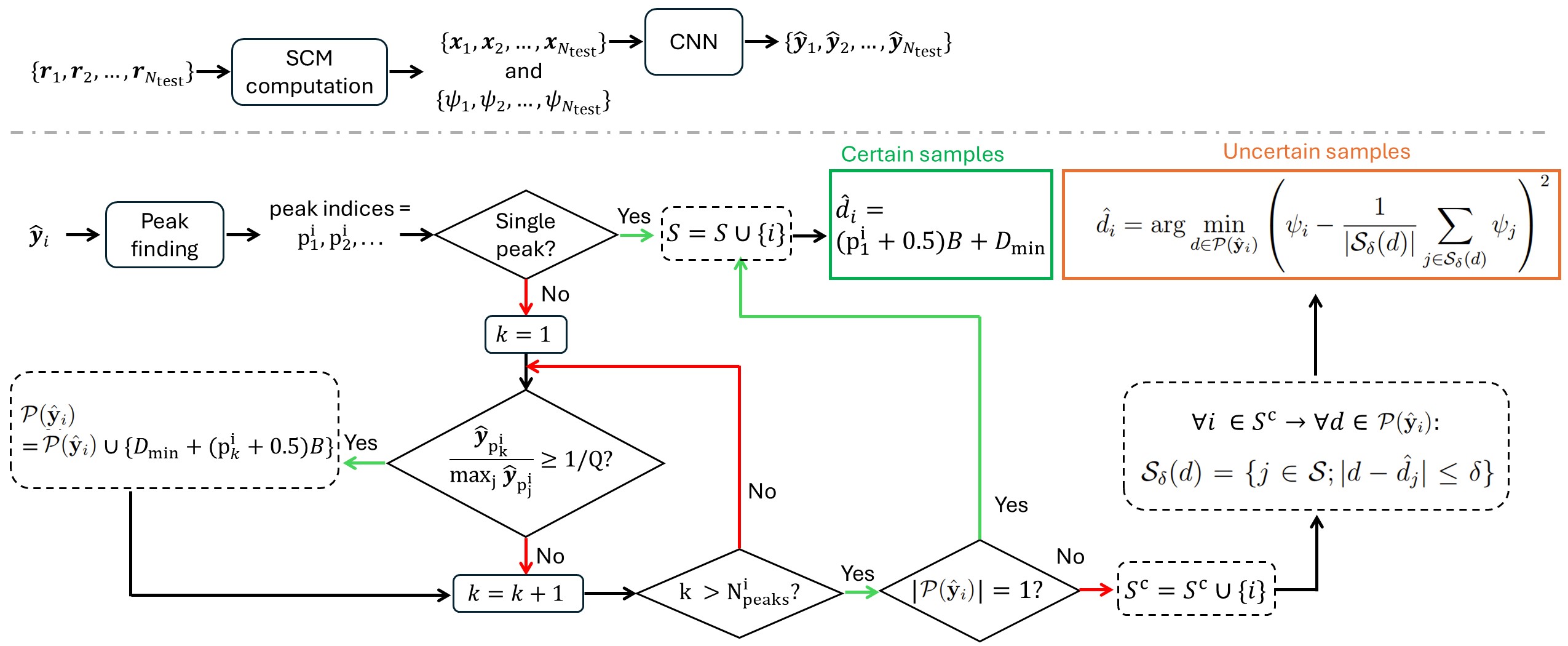}
        \caption{\added[]{A flowchart for JSEA. There are two major steps in JSEA: First, the CNN model estimates the labels; then the set of certain samples $\sss$ is determined. For the samples in $\sss$, the CNN estimates are considered the final range estimates. Finally, for each of the uncertain samples in $\sss^c$, the final range estimate is selected from the set of peak locations, according to \eqref{eq:uncertain_pseudo}.}}
        \label{fig:JSEA-flowchart}
\end{figure*}

For each sample in $\Dtest$, the pre-trained classification model generates $\hat{\mathbf{y}}_i = \hat{p}(d | \tilde{\mathbf{C}})$. After partitioning the samples into $\sss$ and $\sss^c$ based on their peakwise uncertainty, the final range \replaced[]{estimate $\hat{d}_i$}{$\hat{d}_i$ estimate} for $\mathbf{x}_i \in \sss$ is the distance corresponding to the only peak of $\hat{\mathbf{y}}_i$. For $\mathbf{x}_i \in \sss^c$, let $\mathcal{P}(\hat{\mathbf{y}}_i)$ be the set of distances corresponding to all significant peak locations of $\hat{\mathbf{y}}_i$. The goal is to determine a member of $\mathcal{P}(\hat{\mathbf{y}}_i)$ as the pseudo-label for $\mathbf{x}_i$ by exploiting 
\begin{equation}\label{eq:max_p_d_psi}
\hat{d}_i = \arg \max_{d \in \mathcal{P}(\hat{\mathbf{y}}_i)} p(d | \psi = \psi_i).
\end{equation}
In a homogeneous boundary-less acoustic medium, the received power (on a single hydrophone) from a point source at distance $d$ is proportional to $1/d$. In ocean waveguides, however, the received power fluctuates around a decreasing function $\psi_0(d)$; the fluctuations are affected by the ocean variability and ambient noise as well, leading to the low resolution of the estimates based on $p(d | \psi)$. Assuming that $\psi(d)$ is normally distributed with the mean $\psi_0(d)$ and considering a non-informative prior distribution on $\psi$, \eqref{eq:max_p_d_psi} reduces to
\begin{equation}
    \hat{d}_i = \arg \min_{d \in \mathcal{P}(\hat{\mathbf{y}}_i)} (\psi_i - \psi_0(d))^2.
\end{equation}
Since $\psi_0$ is unknown, we approximate $\psi_0(d)$ for each $d \in \mathcal{P}(\hat{\mathbf{y}}_i)$ using the samples in $\sss$ with \cite{kari2024joint} $\hat{\psi}_{0}(d) = \frac{1}{|\sss_{\delta}(d)|} \sum_{j \in \sss_{\delta}(d)} \psi_j$, where $\sss_{\delta}(d) = \{j \in \sss; |d-\hat{d}_j| \leq \delta \}$, and $\psi_j$ is the received power of the $j$-th sample. Therefore,
\begin{equation}\label{eq:uncertain_pseudo}
    \hat{d}_i = \arg \min_{d \in \mathcal{P}(\hat{\mathbf{y}}_i)} \left(\psi_i - \frac{1}{|\sss_{\delta}(d)|} \sum_{j \in \sss_{\delta}(d)} \psi_j \right)^2.
\end{equation}
By defining the pseudo-labels for $\mathbf{x}_i$ as $\mathbf{y}_i^{\mathsf{pseudo}} \triangleq \eta(\hat{d}_i)$, we have achieved the range estimates for all given test samples. If the model is to be used for another source in the same environment, one should also adapt the feature extractor by minimizing the JSEA loss defined by $\mathcal{L}_{\text{JSEA}} = \sum_{i=1}^{\Ntest} \mathsf{CE}(\mathbf{y}_i^{\mathsf{pseudo}}, \hat{\mathbf{y}}_i)$. \added[]{Fig. \ref{fig:JSEA-flowchart} provides a flowchart for the JSEA algorithm.}

\added[]{While SHOT defines the pseudo-labels only for the samples in $\sss$, JSEA defines pseudo-labels for all test samples. In both SHOT and JSEA, pseudo-labels for certain samples are the final range estimates. However, to obtain the final range estimates for uncertain samples, SHOT relies on re-training the network for a few iterations using the pseudo-labeled certain samples, while JSEA relies on the peak selection process based on the received signal power.}

\section{\replaced[]{}{Empirical }Evaluation} \label{sec:simulations}
\subsection{Synthetic data}
\begin{figure}[ht]
\centering
    \includegraphics[width=\linewidth]{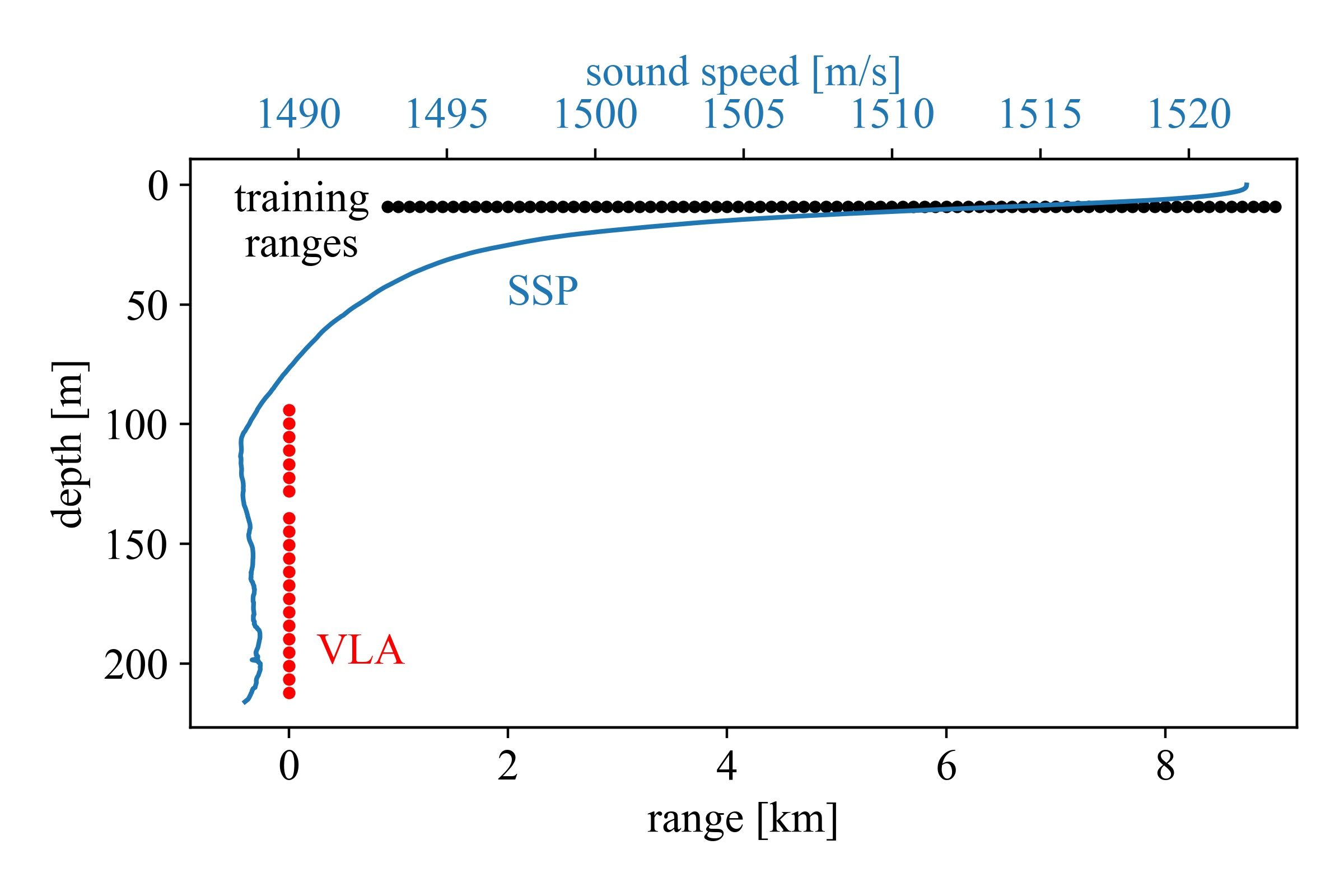}
    \caption{KRAKEN environment based on the SWellEx-96 experiment. The receiver array is a 21-element array depicted in red. The uniformly sampled source ranges that are used for training are shown with black dots. Also the SSP, which is the average SSP from the SWellEx-96 experiment, is shown with the blue graph.}
    \label{fig:swellex}
\end{figure}

We first consider data generated using KRAKEN \cite{porter1992kraken}, where the environmental parameters are set according to the SWellEx-96 \cite{gemba2017adaptive,wang2018underwater} experiment. As depicted in Fig. \ref{fig:swellex}, the training data consist of source ranges between $0.85$~km and $9.05$~km with $10$~m increments (thus, $821$ distinct samples), and a $21$-element vertical linear array placed at depths from $94.125$~m to $212.25$~m to mimic the SWellEx-96 data. The source depth is assumed to be $9$~m and transmits a monotone signal at a frequency of $f=109$~Hz. We consider a flat bathymetry at depth $216.5$~m in our simulations and used the average SSP from the SWellEx-96 measurements. The bottom constitutes of $3$ layers whose depths, densities, and attenuation match the SWellEx-96 environment. \par

To generate a test environment that has a sound-speed $c(z)$ at depth $z$, which is mismatched with that of the training environment, $c_0(z)$, we perturb $c_0(z)$ using a constant gradient difference of $\Delta c$~(m/s)/m, represented as $c(z) = c_0(z) + \frac{\Delta c}{216.5} (z - 216.5)$, where $0 \leq z \leq 216.5$~m. We generate data for $500$ random source ranges between $D_{\min} = 900$~m and $D_{\max} = 9$~km, and divide the region into $100$~m-long intervals, each represented by its mid point. Consequently, the labels and classifier outputs are 82-dimensional (i.e., $M=82$). In addition, to emulate the real ocean ambient noise during test time, we add a randomly selected noise segment $w$ from the KAM11 experiment \cite{tomasi2011predictability} (scaled to have a desired signal-to-noise ratio in the data) to the generated signal for each hydrophone. Since the proposed method takes a batch of test data, we use the average array SNR of the batch, defined as
\[\text{SNR [dB]} = 10 \log_{10} \frac{\sum_{i=1}^{\Ntest}\sum_{l=1}^L |r_{i,l}|^2}{\Ntest L \psi_w},\]
where $|r_{i,l}|^2$ is the received signal power at the $l$-th hydrophone of the $i$-th sample and $\psi_w$ is the noise power spectral density at the frequency $f=109$ Hz.\par
We evaluate models using MAE and probability of credible localization (PCL) \cite{wang2019deep} defined as
\begin{align*}
\text{PCL}(\zeta) & = \frac{100}{\Ntest} \sum_{i=1}^{\Ntest} \mathbf{1}_{|d_i - \hat{d}_i| \leq \zeta d_i}, \\ 
\text{MAE} & = \frac 1 {\Ntest} \sum_{i=1}^{\Ntest} |d_i - \hat{d}_i|,
\end{align*}
where $d_i$ and $\hat{d}_i$ are respectively the true and estimated distances of the $i$-th sample. Probability of credible localization can be interpreted as the probability that the range estimate lies within a certain percentage bound of the true range, defined here as $\zeta = 10\%$. 

The methods evaluated include:
\begin{itemize}
    \item O-MFP (Oracle-MFP): The MFP approach tuned to the test environment. This method is presented just for reference and is expected to outperform all other methods because it has access to the test environment parameters.
    \item M-MFP (mismatched MFP): The MFP approach tuned to the training environment. This method \replaced[]{is assumed to have}{, in contrast with data-driven methods, has} access to the training data.
    \item CNN-c: A CNN-based model for classification\replaced[]{, trained once on the training set, and not adapted to account for mismatch}{}.
    \item SHOT: The CNN-c equipped with the adaptation mechanism in \eqref{eq:SHOT_loss}, with $\mu_{\text{SHOT}} = 5\times10^{-6}$ and $\beta = 1$.
    \item JSEA-c:  The CNN-c equipped with the pseudo-label estimator in \eqref{eq:uncertain_pseudo}, where $\delta=500$ m.
    \item CNN-r: A CNN-based model for regression.
    \item JSEA-r: The CNN-r followed by an MC-dropout mechanism to generate output PMFs and equipped with the pseudo-label estimator in \eqref{eq:uncertain_pseudo}, where $\delta=500$ m.
\end{itemize}

To select the label smoothing hyperparameter $\sigma$, we compared the performance of the trained model on the validation data for different values of $\sigma$ and found that $\sigma=2$ works well. We also use a learning rate of $10^{-4}$ and a batch size of $128$ for the Adam optimizer. We randomly split the set of uniformly sampled ranges into $82 \%$ training and $18 \%$ validation sets. To prevent over-fitting and converge better to the minimum, we adopt a decreasing learning rate - we multiply the learning rate by $0.1$ whenever there is no reduction in the validation error after $75$ iterations, and we stop training if there is no reduction in the validation error after $125$ iterations \cite{chen2021model}. To enhance robustness against noise, after training the network with clean training data, we continue training with noisy data, where the inputs are contaminated by white Gaussian noise with an SNR in $\{2, 4, 6, 8, 10, 12, 14, 16\}$ dB. The test results are averaged over $100$ noise realizations.\par

\begin{figure}[htb]
\begin{subfigure}[t]{0.45\textwidth}
\centering
    \includegraphics[width= \linewidth]{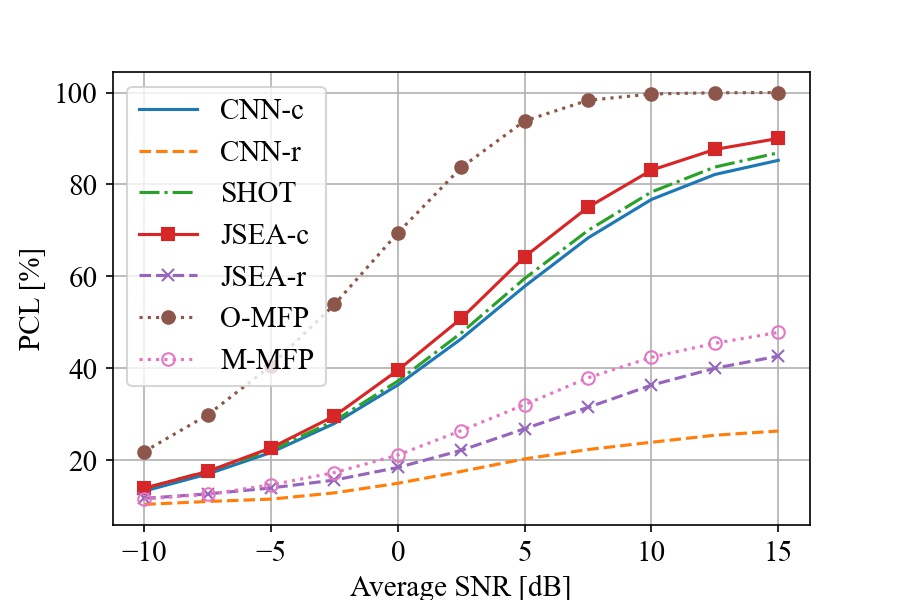}
    \caption{PCL(10\%)}
    \label{fig:mae_snr}
\end{subfigure}
~
\begin{subfigure}[t]{0.45\textwidth}
\centering
    \includegraphics[width= \linewidth]{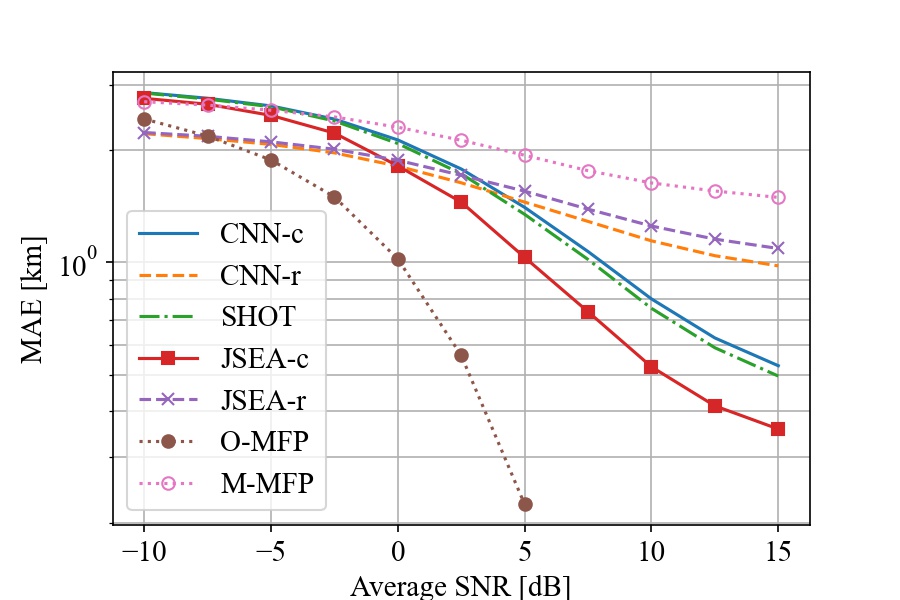}
    \caption{MAE}
    \label{fig:pcl_snr}
\end{subfigure}
\caption{Performance versus SNR using noise data from KAM11 experiment and with $\Delta D = 4$ m for the shallow source with the monotone signal at $f=109$ Hz, evaluated with metrics: (a) PCL and (b) MAE. While the O-MFP provides a performance bound as it has access to training data generated in an environment matched to the test environment, the M-MFP shows the performance degradation of the MFP method when provided with the mismatched data.}
\label{fig:snr_performance}
\end{figure}

Fig. \ref{fig:snr_performance} shows the superiority of the JSEA-c and \replaced[]{CNN-based classification}{CNN-c} methods over M-MFP when the training and testing environments have a depth mismatch of $\Delta D = 4$ m (i.e, the ocean depth in the test environment is $220.5$ m), and there is a mismatch in the noise statistics of the test data compared to the training data as well. The proposed JSEA-c outperforms other methods including SHOT, however, its improvement highly depends on the success of the pre-trained model in generating enough certain labels. Although, JSEA-r can successfully improve the regression results as seen in Fig. \ref{fig:snr_performance}, we will not include the regression models in the following experiments due to their inferior performance\replaced[]{ compared to the classifier-based approaches}{}. In this scenario, the M-MFP shows inferior performance with respect to classification methods, in the presence of depth mismatch.\par

\begin{figure}[htb]
\begin{subfigure}[t]{0.45\textwidth}
\centering
    \includegraphics[width= \linewidth]{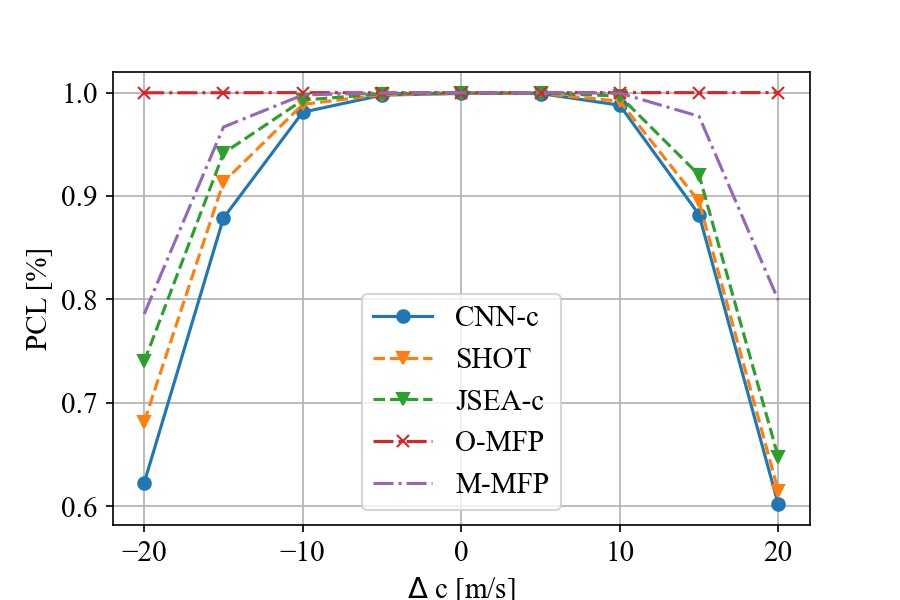}
    \caption{PCL(10\%) under mismatch in SSP.}
    \label{fig:mae_ssp_tiny}
\end{subfigure}
~
\begin{subfigure}[t]{0.45\textwidth}
\centering
    \includegraphics[width= \linewidth]{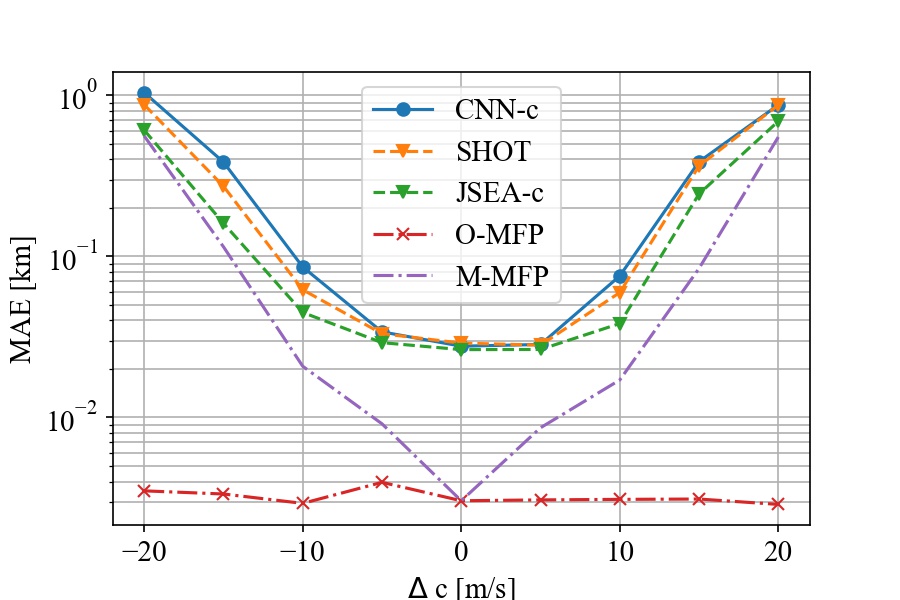}
    \caption{MAE under mismatch in SSP.}
    \label{fig:pcl_ssp_tiny}
\end{subfigure}
\caption{Performance against variation in SSP mismatch at an SNR of $15$ dB.}
\label{fig:ssp_performance}
\end{figure}

Fig. \ref{fig:ssp_performance} shows the performance comparison of the methods with variation in the degree of SSP mismatch. In the current scenario, mismatch in SSP gradient is assessed, by varying the gradient difference $\Delta c$ between training and test data. As seen in Fig. \ref{fig:ssp_performance}, the M-MFP outperforms the DL models; however, it benefits from using training data during inference, which is not available to the other methods. While SHOT can improve the CNN-c results, JSEA-c has the best performance among the DL methods.\par

\begin{figure}[htb]
\begin{subfigure}[t]{0.45\textwidth}
\centering
    \includegraphics[width= \linewidth]{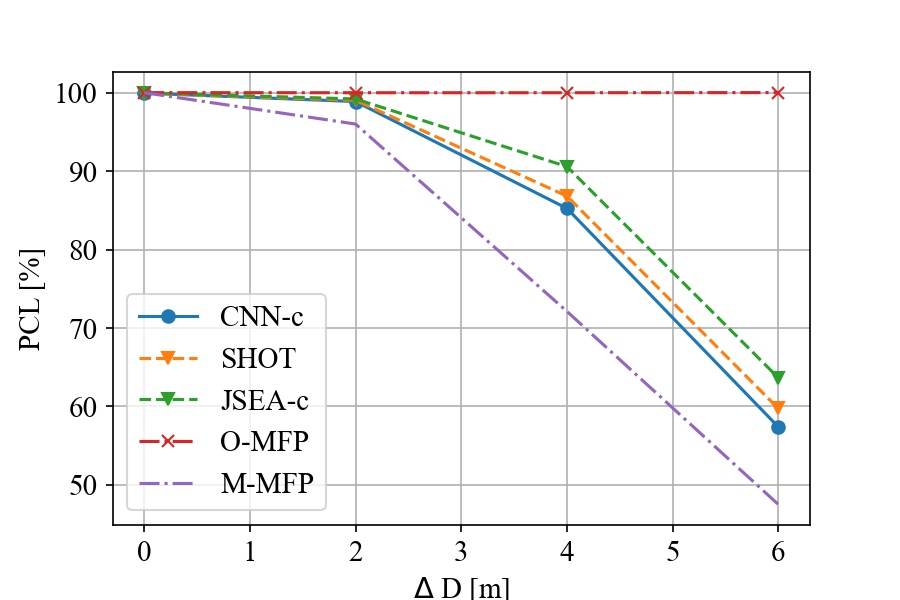}
    \caption{PCL(10\%) under mismatch in \replaced{depth}{SSP}.}
    \label{fig:pcl_ssp}
\end{subfigure}
~
\begin{subfigure}[t]{0.45\textwidth}
\centering
    \includegraphics[width= \linewidth]{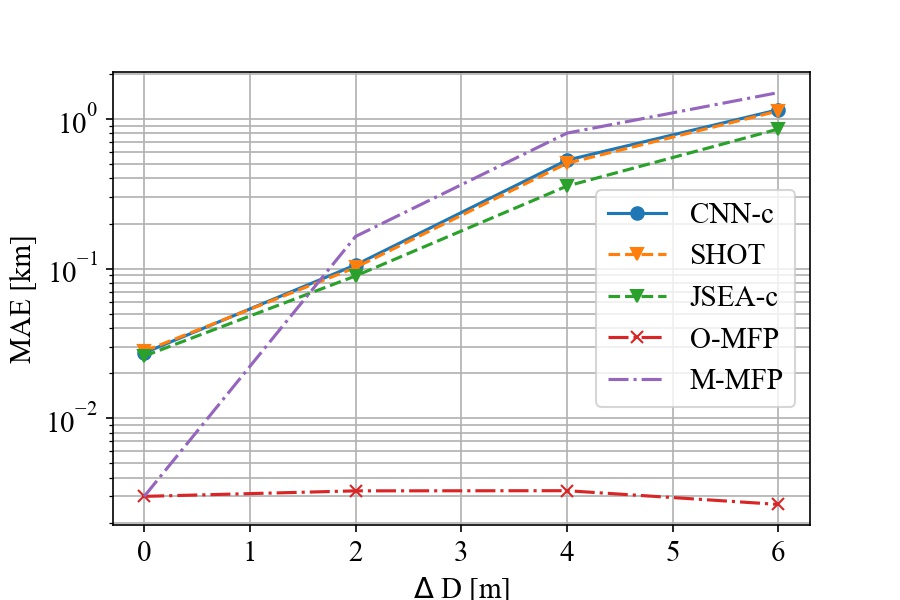}
    \caption{MAE under mismatch in \replaced{depth}{SSP}.}
    \label{fig:mae_ssp}
\end{subfigure}
\caption{Performance against variation in magnitude of depth mismatch at an SNR of $15$ dB.}
\label{fig:depth_performance}
\end{figure}

Next we consider the depth mismatch and evaluate the performance of the algorithms for $\Delta D \in \{0, 2, 4, 6\}$ m/s. As seen in Fig. \ref{fig:depth_performance}, when the test and training environments match (i.e., $\Delta D = 0$), the MFP (in this case, the M-MFP is not mismatched, hence it is the same as O-MFP) performs the best. However, depth mismatches lead to the M-MFP's performance \replaced[]{deteriorating}{deterioration}, whereas the performance of the CNN-based methods is more robust to this mismatch. Thus, using CNN methods can impart some robustness to localization performance under depth mismatch.\par

\begin{figure}[htb]
\centering
\begin{subfigure}[t]{0.4\textwidth}
\centering
    \includegraphics[width= \linewidth]{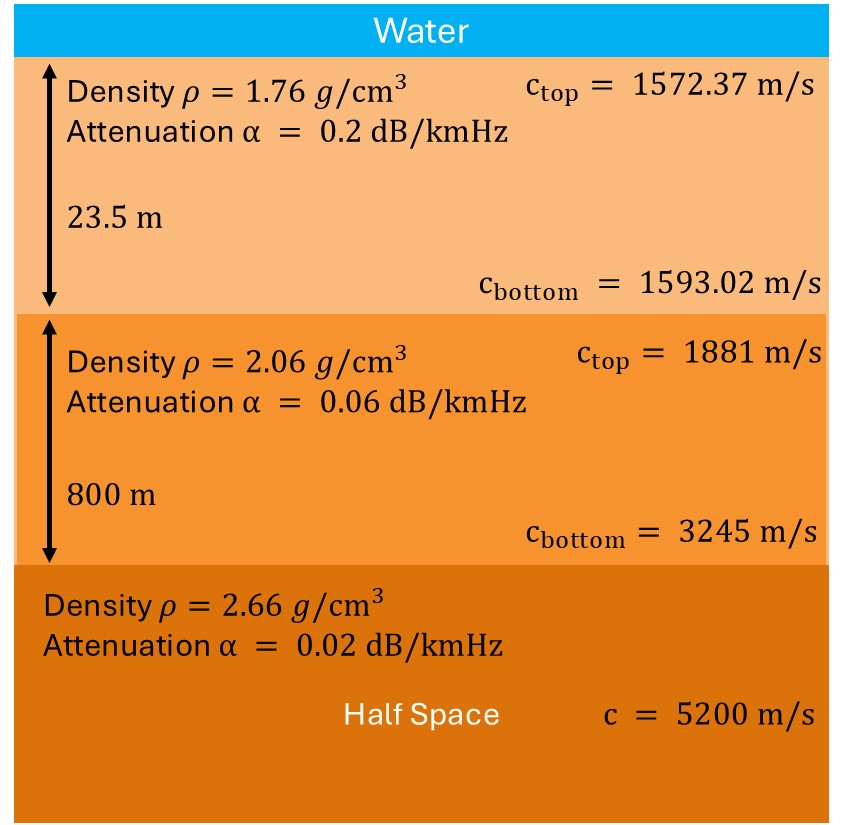}
    \caption{\added[]{Bathymetry of the training environment, inspired by the SWellEx-96 environment.}}
    \label{fig:swellex_sediment_bathymetry}
\end{subfigure}
~
\begin{subfigure}[t]{0.4\textwidth}
\centering
    \includegraphics[width= \linewidth]{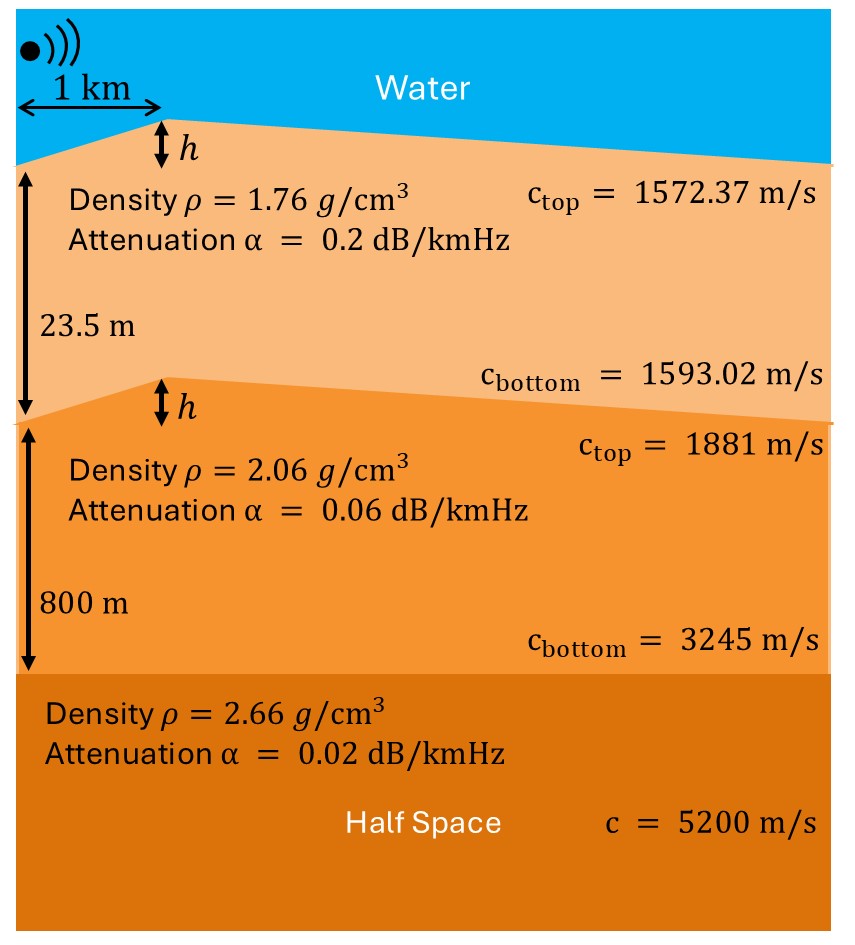}
    \caption{\added[]{Mismatched bathymetry used for testing. There is a seamount structure at 1 km distance from the source, with a height of $h$ meters.}}
    \label{fig:mismatched_sediment_bathymetry}
\end{subfigure}
\caption{\added[]{Bathymetry mismatch used to evaluate the performance of the JSEA method.}}
\label{fig:bathymetry_mismatch}
\end{figure}

\begin{figure}[htb]
\begin{subfigure}[t]{0.45\textwidth}
\centering
    \includegraphics[width= \linewidth]{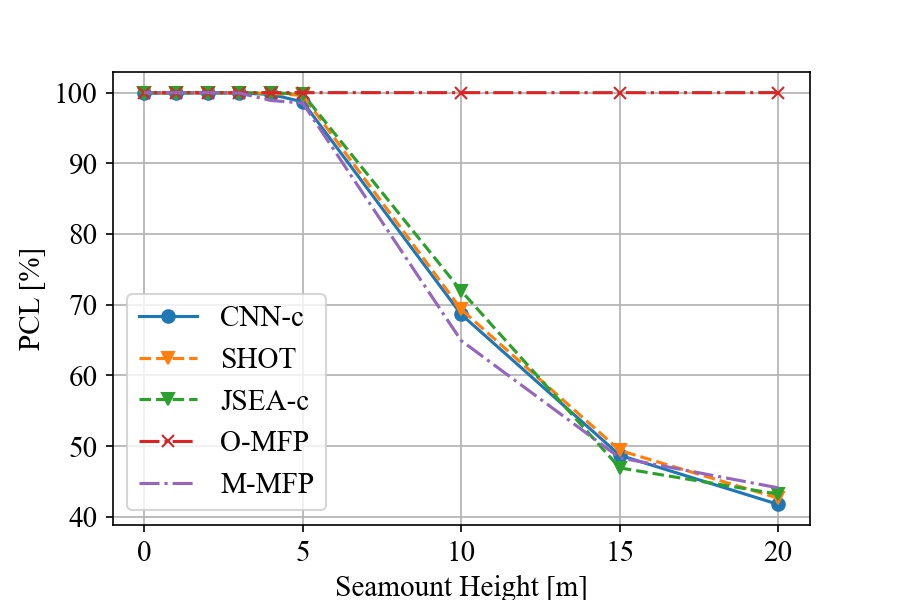}
    \caption{\added[]{PCL(10\%) under mismatch in bathymetry.}}
    \label{fig:mae_bathy}
\end{subfigure}
~
\begin{subfigure}[t]{0.45\textwidth}
\centering
    \includegraphics[width= \linewidth]{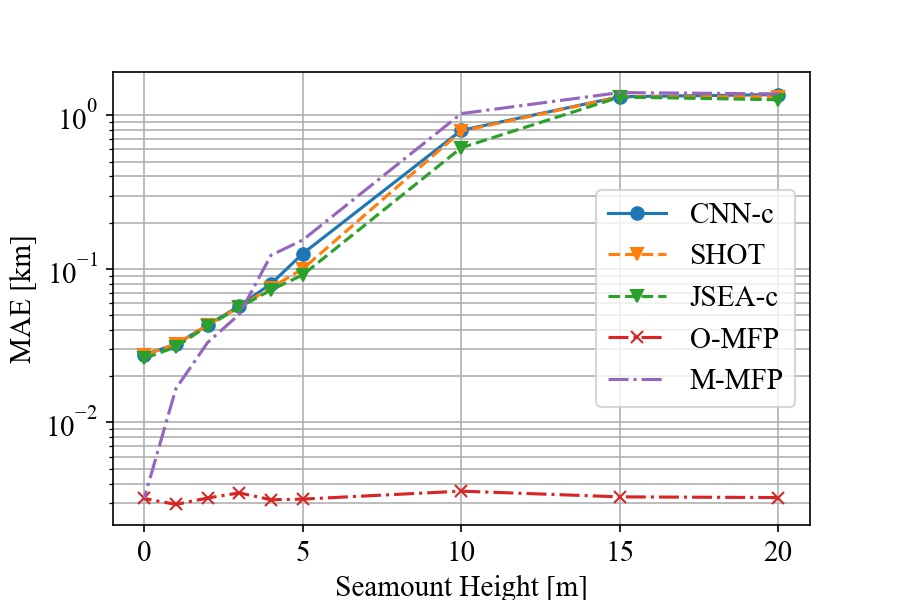}
    \caption{\added[]{MAE under mismatch in bathymetry.}}
    \label{fig:pcl_bathy}
\end{subfigure}
\caption{\added[]{Performance against variation in bathymetry at an SNR of $15$ dB.}}
\label{fig:bathymetry_performance}
\end{figure}

\added[]{To further investigate the effect of environmental mismatches, we evaluate the performance against mismatches in bathymetric shape/fluctuations, as shown in Fig. \ref{fig:bathymetry_mismatch}. While the training environment includes a flat bathymetry, the test environment contains a seamount. Fig. \ref{fig:bathymetry_performance} shows the MAE and PCL results versus the seamount height. The results in Fig. \ref{fig:bathymetry_performance} verify the superior performance of the JSEA-c in terms of both MAE and PCL for this scenario.}

\begin{figure}[htb]
\begin{subfigure}[t]{0.45\textwidth}
\centering
    \includegraphics[width= \linewidth]{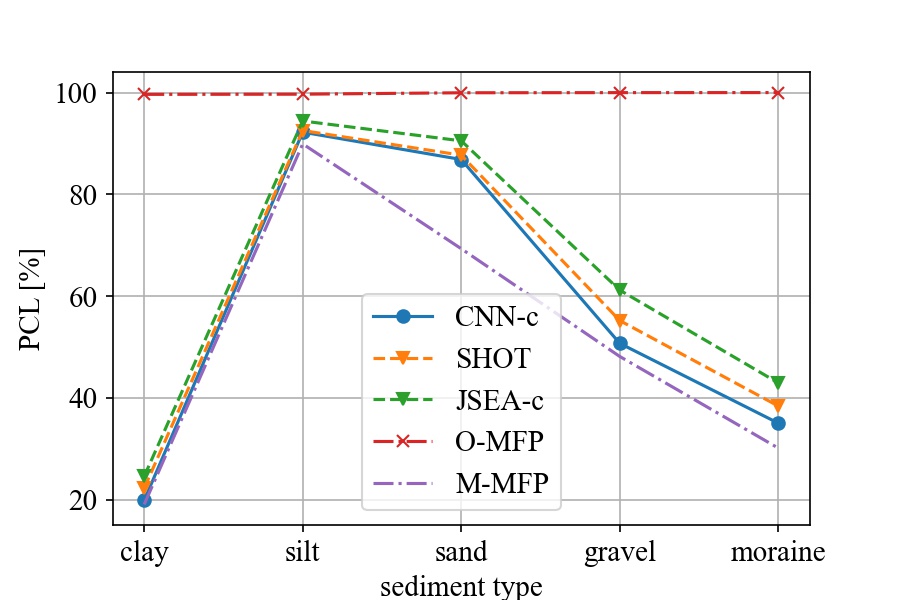}
    \caption{\added[]{PCL(10\%) under mismatch in sediment type.}}
    \label{fig:mae_depth}
\end{subfigure}
~
\begin{subfigure}[t]{0.45\textwidth}
\centering
    \includegraphics[width= \linewidth]{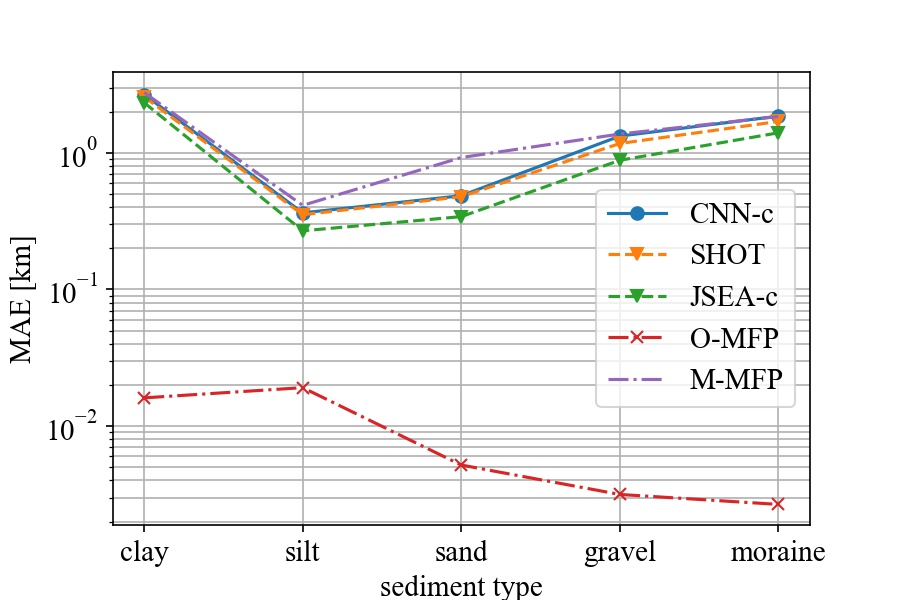}
    \caption{\added[]{MAE under mismatch in sediment type.}}
    \label{fig:pcl_depth}
\end{subfigure}
\caption{\added[]{Performance against mismatch with various sediment types at an SNR of $15$ dB.}}
\label{fig:sediment_performance}
\end{figure}

\added[]{Mismatch in bottom sediment type is another type of environmental mismatch that commonly arises in localization problems. Each sediment type varies in density, sound speed, and attenuation coefficient, which affect the acoustic wave propagation through bottom interactions. For the sake of demonstrating robustness to mismatch, we consider only fluid sediment types here, and do not consider shear speed or attenuation. To evaluate the performance against mismatches in sediment type, we generate training data according to the sediment type provided in Fig.\ref{fig:swellex_sediment_bathymetry} and evaluate the performance on test data generated in environments with sediment types defined according to Table \ref{tab:sediment_types} \cite{jensen2011computational}. Fig.  \ref{fig:sediment_performance} shows the MAE and PCL results versus the sediment type. These results verify the superior performance of the JSEA-c in terms of both MAE and PCL for this scenario.}

\begin{table*}[h]
    \centering
    \caption{\added[]{Different sediment types \cite{jensen2011computational}}}
    \begin{tabular}{|c|c||c|c|c|c|c|}
    \hline
          & training sediment & clay & silt & sand & gravel & moraine \\
         \hline \hline
         density $\rho \;\;  [g / cm^3]$ & 1.76 & 1.5 & 1.7 & 1.9 & 2 & 2.1 \\
         \hline 
         minimum sound speed $c_{\text{min}} [m/s]$ & 1572.37 & 1500 & 1575 & 1650 & 1800 & 1950 \\
         \hline
         maximum sound speed $c_{\text{max}} [m/s]$ & 1593.02 & 1520 & 1595 & 1670 & 1820 & 1970 \\
         \hline
         attenuation $\alpha$ [dB/ (km Hz)] & 0.2 & 0.2 & 1 & 0.8 & 0.6 & 0.4 \\
         \hline
         
    \end{tabular}
    \label{tab:sediment_types}
\end{table*}

Overall, we observe that using TTA can enhance the performance of the CNN-based approach by allowing better adaptation to the test environment and introducing more robustness to mismatch. This is demonstrated by the improvement in SHOT and JSEA in comparison to CNN. Using the received signal power information further improves the performance, as seen in the improvement in JSEA as compared to SHOT.

\subsection{SWellEx-96 Data}\label{sec:SWellEx}

\begin{figure}
    \centering
    \includegraphics[width=\linewidth]{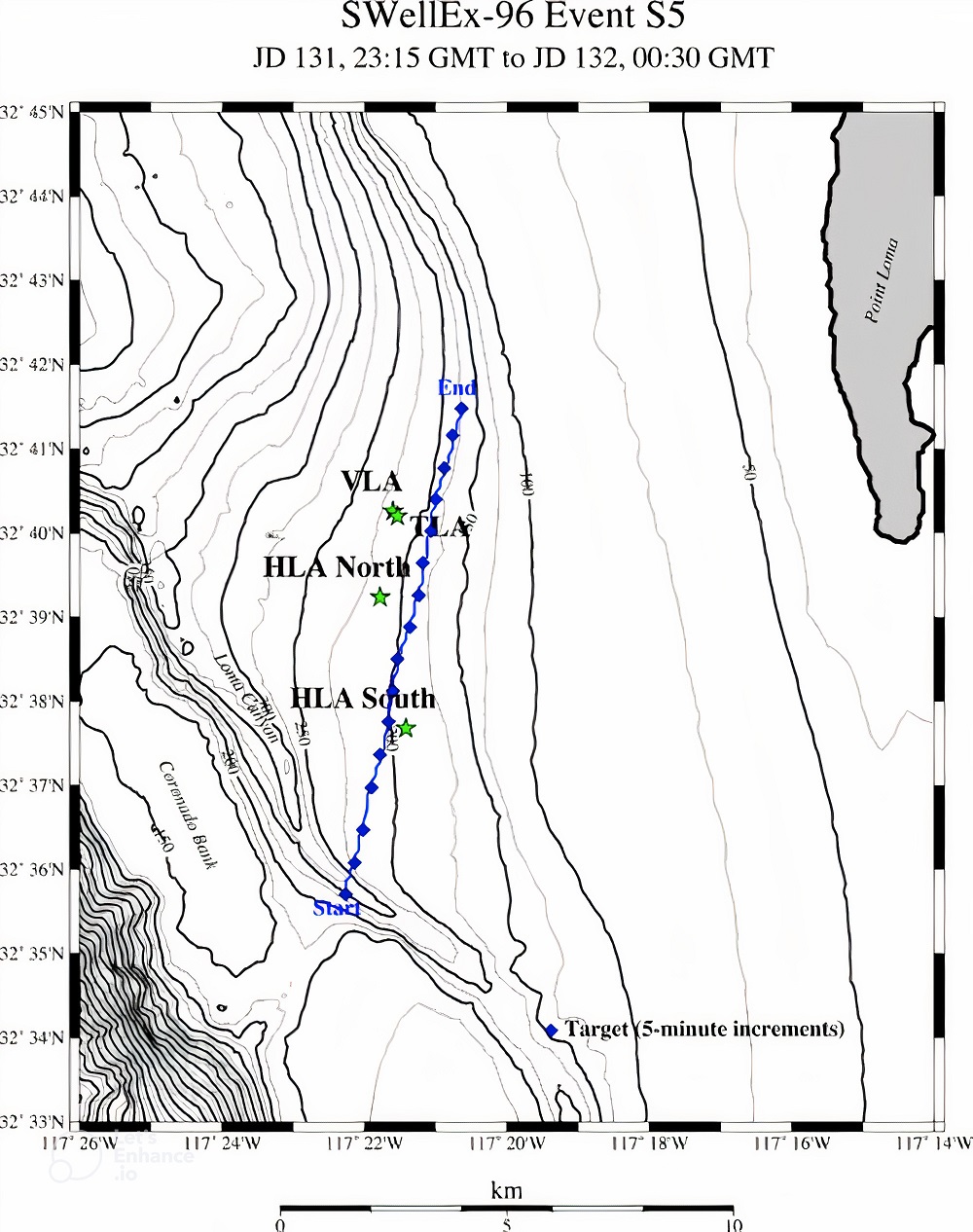}
    \caption{The ship track for the SWellEx-96 S5 event \cite{Booth2015SWellex}. \replaced[]{The localization problem considered involves estimating}{We aim to estimate} the distance of the ship to the VLA receiver.}
    \label{fig:S5track}
\end{figure}

We use data from the event S5 of the public dataset SWellEx-96 experiment \cite{Booth2015SWellex, hursky2001matched, orris2000matched}. In this event, two sources, a shallow source at a depth of $9$ m and a deep source at a depth of $54$ m, were towed by a ship moving in a straight line as shown in Fig. \ref{fig:S5track}. The whole event took $75$ minutes, during which each source transmitted a set of frequencies with a specific power level, recorded by several arrays at the sampling frequency of $F_s = 1500$~Hz. We use the recordings from the $21$-element ($L=21$) vertical line array (VLA) and provide the results for the shallow source.

To obtain the Fourier transform and the corresponding SCM $\mathbf{C}$ used for localization, we use a $3$-second-long signal divided into $P=5$ segments, each of length $1$ second and a $50\%$ overlap, tapered by a Kaiser window with $\beta=9.24$. We obtain the true source range by linear interpolation between the GPS recordings provided\replaced[]{ assuming the ship followed a straight-line path between GPS updates}{}.

\begin{figure*}[t]
    \centering
    \begin{subfigure}[b]{0.32\textwidth}
        \centering
        \includegraphics[width=\textwidth]{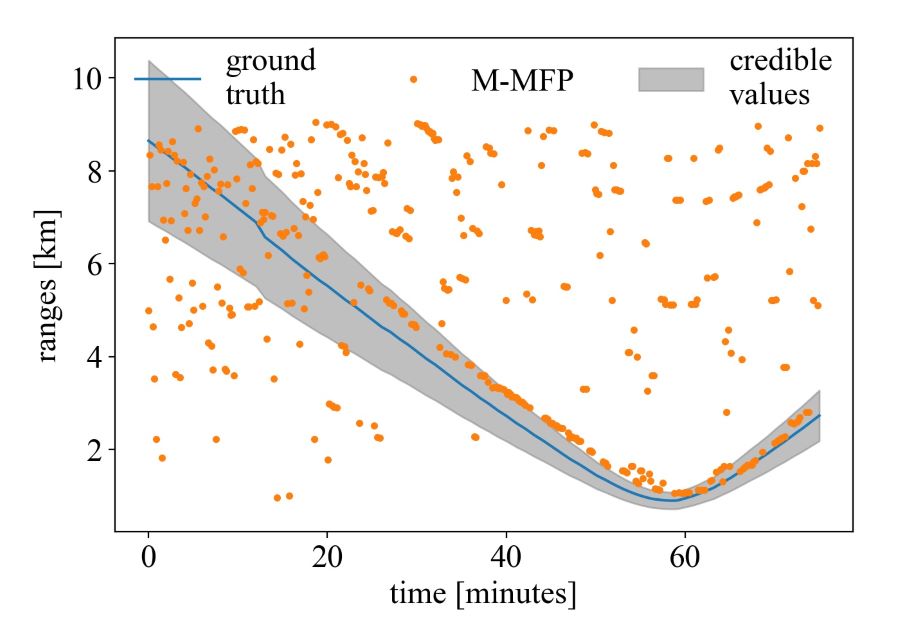}
        \caption{M-MFP.}
        \label{fig:SWellEx_MFP}
    \end{subfigure}
    \hfill
    \begin{subfigure}[b]{0.32\textwidth}
        \centering
        \includegraphics[width=\textwidth]{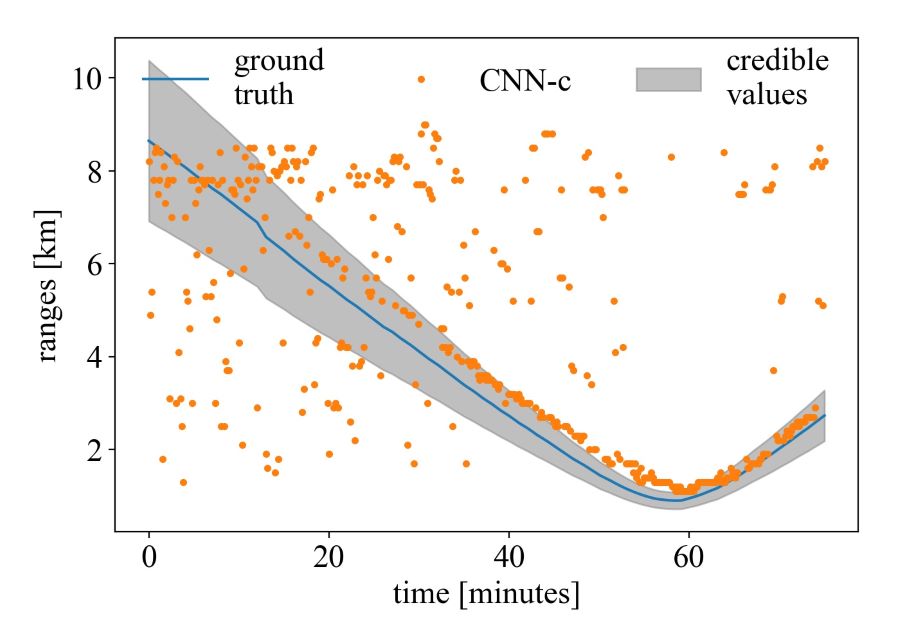}
        \caption{CNN-c.}
        \label{fig:SWellEx_CNN}
    \end{subfigure}
    \hfill
    \begin{subfigure}[b]{0.32\textwidth}
        \centering
        \includegraphics[width=\textwidth]{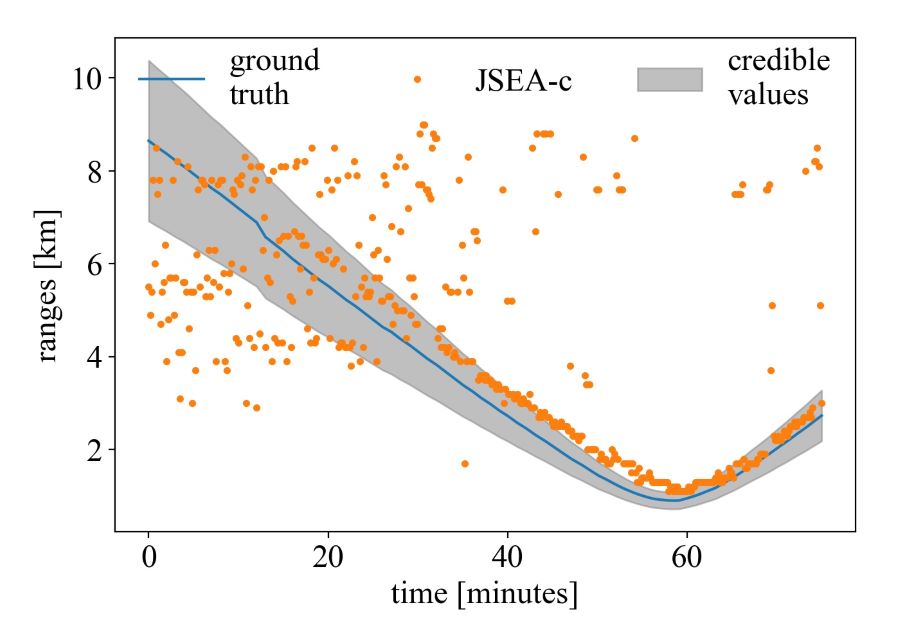}
        \caption{JSEA-c.}
        \label{fig:SWellEx_JSEA}
    \end{subfigure}
    \caption{Localization results for SWellEx-96 data. The blue curve indicates the ground-truth values of range at each time. The orange dots in each plot, indicate the corresponding estimated ranges. The gray cone around the blue curve indicates an interval from $0.9$ to $1.1$ of the true range. Observe how with JSEA, the predicted ranges are concentrated closer to the true ranges.}
    \label{fig:SWellEx}
\end{figure*}

Fig. \ref{fig:SWellEx} shows the ranging results of M-MFP, CNN-c, and JSEA-c on the real-world narrowband data at frequency $f=109$~Hz. The results in Fig. \ref{fig:SWellEx_MFP} suggest that M-MFP performs poorly in this scenario, as it leans towards overestimating the ranges for the samples around $t = 60$~min which are closest to the VLA and have a relatively higher SNR. This problem is absent in the CNN-c approach as observed in Fig. \ref{fig:SWellEx_CNN}. Nevertheless, both M-MFP and CNN-c tend to deviate significantly from the true ranges in the low SNR regions, i.e., larger source ranges ($t \leq 40$ min). \replaced[]{While}{Whereas} this problem is still present in JSEA-c, it is considerably less severe, yielding an overall better MAE and PCL, as shown in Table \ref{tab:swellex_results}.\par

To understand the effect of the JSEA-c approach on reducing the error, we show the partitioning of the samples into certain and uncertain sets based on the PU scores, in Fig. \ref{fig:PU}. This figure denotes that most of the uncertain samples are in the low SNR region, where the M-MFP and CNN-c fail. This means that the JSEA-c approach relies on the certain samples (mainly the correctly estimated samples which tend to be from the high SNR region) to rectify the incorrect outputs of the CNN-c. This might also be, in part, due to the fact the samples around $t=60$~min are not only higher in SNR, but also their environment is matched the most to the training environment, as the depth we use for training environment is $216.5$~m, which is the depth at the VLA location.

\begin{table}
    \centering
    \caption{Localization performance for SWellEx-96 data.}
    \begin{tabular}{|c||c|c|c|}
    \hline
          & M-MFP & CNN-c & JSEA-c\\
         \hline \hline
         MAE [km]  & 2.387 & 1.746 & 1.430 \\
         \hline 
         PCL ($20\%$) & 31.78 & 38.89 & 42.67\\
         \hline
         
    \end{tabular}
    \label{tab:swellex_results}
\end{table}

\begin{figure*}[htb]
\centering
\begin{subfigure}[t]{0.4\textwidth}
\centering
    \includegraphics[width= \linewidth]{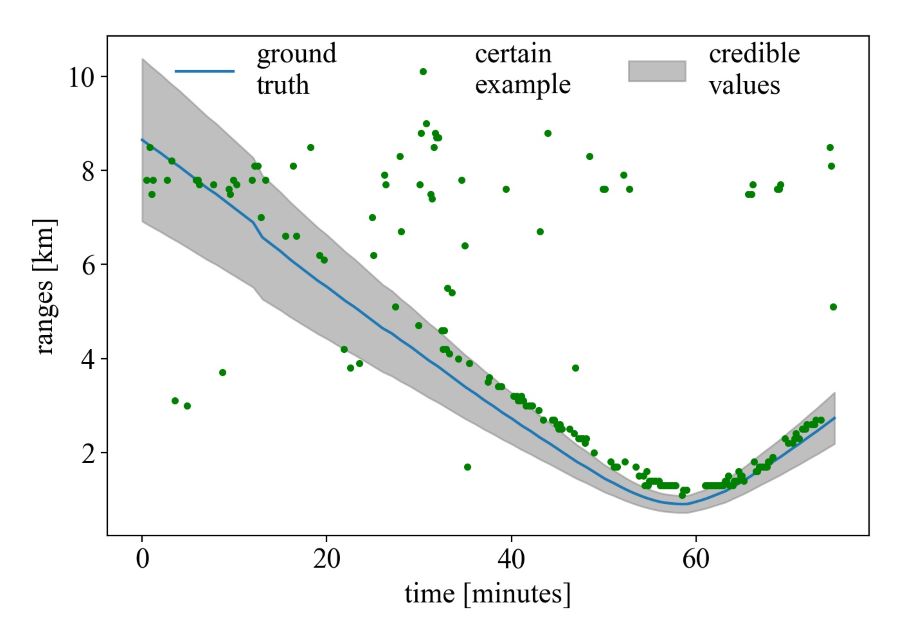}
    \caption{Certain examples.}
    \label{fig:sweelex_certain}
\end{subfigure}
~
\begin{subfigure}[t]{0.4\textwidth}
\centering
    \includegraphics[width= \linewidth]{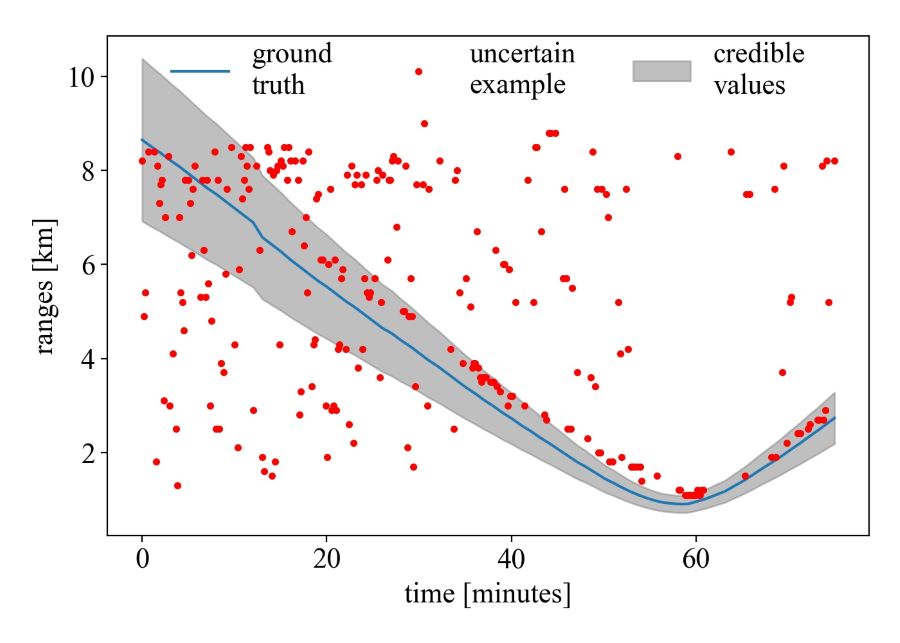}
    \caption{Uncertain examples.}
    \label{fig:swellex_uncertain}
\end{subfigure}
\caption{Partitioning of the CNN-c predictions for SWellEx-96 data into certain and uncertain examples using PU metric. The solid blue line indicates the ground-truth. The left plot shows certain examples and the right plot shows uncertain examples.}
\label{fig:swellex_certain_uncertain}
\end{figure*}

\added[]{\protect \section{Discussion, Extensions, And Limitations of JSEA}\label{sec:ext-lim}}
\added[]{Whereas the empirical evaluations in the previous section verify the efficacy of JSEA in several different scenarios, it is a sophisticated method with some inherent limitations. This section addresses hyperparameter selection, JSEA limitations and possible algorithmic extensions, and complexity comparison with other methods.}

\added[]{\protect \subsection{Hyperparameter Selection}}
\added[]{Successful operation of JSEA relies on reasonable selection of the hyper-parameters. The hyper-parameters $\sigma$ and $B$ are related to converting the regression problem to a classification one, and can be selected by evaluating the performance on a validation set, and adjusting their values according to the following guidelines. Moreover, the adaptation hyper-parameters $\delta$ and $Q$ can be determined based on the desired performance and any prior knowledge on the degree of the mismatch between training and test environments, according to the following discussion. In the following, we explain the effect of each of these four hyper-parameters.}

\begin{itemize}
    \item \added[]{$\sigma$: A large $\sigma$ leads to ground-truth labels that are heavy-tailed and closer to a uniform distribution and cannot indicate the distinction between different classes clearly. This in turn slows the training due to weak supervisory signal and can lead to an inferior accuracy. On the other hand, a small $\sigma$ leads to ground-truth labels that are light-tailed and closer to a point mass (spike), rendering labels one-hot-coded, which can exacerbate the overfitting of the model by removing the distance information between different classes. Hence the $\sigma$ value is chosen to balance the tradeoff between these two factors.}

    \item \added[]{$B$: A larger $B$ leads to lower precision (a coarser quantization) in localization which bounds the best localization performance possible, but also yields a less complex model with fewer output classes which makes classification easier and enhances performance. So in practice, $B$ should be determined according to this tradeoff between the precision required and the desired model complexity. Based on this tradeoff, and also following Chen and Schmidt [1], who used a similar CNN-based architecture for source ranging using the SWellEx-96 dataset, we select $B = 100$ m.}
    
    \item \added[]{$Q$: The hyperparameter $Q$ adjusts the sensitivity of the algorithm to mismatches; the larger  $Q$ is, the more sensitive JSEA becomes to mismatches. As such, $Q$ should be selected to draw a trade-off between the desired size of $\sss$ and the minimum confidence desired in pseudo-labels selected. Determining this trade-off is naturally tied to $\sigma$ and $\delta$. We empirically observed that $Q=10$ leads to a reasonable trade-off.}
    
    \item \added[]{$\delta$: This hyper-parameter should be small enough to be able to distinguish among the peaks to rectify the labels of uncertain samples, i.e.,
    \[
    \forall j \in \sss^c: \;\; \forall d_1, d_2 \in \mathcal{P}(\hat{\mathbf{y}}_j), \;\; \text{if} \;\; d_1\neq d_2, \; \; \sss_{\delta}(d_1) \neq \sss_{\delta}(d_2),
    \]
    and it should be large enough that
    \[
    \forall j \in \sss^c: \;\; \forall d \in \mathcal{P}(\hat{\mathbf{y}}_j), \;\; \sss_{\delta}(d) \neq \phi.
    \]
    If such a $\delta$ cannot be found for a sample $j$, one can either accept the largest output peak as the final estimate, or declare the sample as an uncertain one that cannot be rectified using JSEA. We have used $\delta = 500$ m for all test samples as this value satisfied these conditions.}
\end{itemize}

\added[]{\protect \subsection{Complexity Comparison}}

\added[]{Underwater applications typically require power- and electronic storage-efficient algorithmic designs. Here, we provide a complexity comparison between different methods that were discussed in the paper. Although the calculations are not based on hardware-optimized algorithms, they provide a reasonable baseline for comparison. The complexity analysis is provided in Appendix \ref{app:complexity}. Table \ref{tab:runtime_complexity_comparison} summarizes the complexity and run-time comparison between different methods, when implemented on an NVIDIA RTX 6000 Ada Generation GPU device using CUDA 12.2.}

\begin{table*}[h]
\centering
\caption{\added[]{Memory and Computational Complexity Comparison}}
{\small
\begin{tabular}{|c|c|c|c|}
\hline
Model & \makecell{Computational\\Complexity} & \makecell{Memory\\Complexity} & Run Time $[ms]$ \\
\hline
MFP & $O(\Ntr L^2)$ & $O(\Ntr L)$ & $0.4666$ \\
\hline
CNN-c & $O(N_{\phi}L^2+MN_{\phi})$ & $O(N_{\phi}L^2+MN_{\phi})$ & $0.0014$ \\
\hline
SHOT & $O(N_{\phi}L^2+MN_{\phi} + N_{\text{itr}}N_{\phi} L^2)$ & $O(\Ntest L+ N_{\phi}L^2+MN_{\phi})$ & $1.5225$ \\
\hline
JSEA-c & $O(N_{\phi}L^2+MN_{\phi} + \Ntest)$ & $O(\Ntest L + N_{\phi}L^2+MN_{\phi})$ & $0.2353$ \\
\hline
\end{tabular}
}
\label{tab:runtime_complexity_comparison}
\end{table*}

\added[]{\protect \subsection{Extensions And Limitations}}
\added[]{The proposed JSEA mechanism relies on access to a sufficient number of test samples as well as the ability of the CNN method to classify a sufficient number of them with certainty. Determining the minimum number of certain samples required for successful operation of JSEA is left for future research. Furthermore, variations of JSEA can be developed that use other assumptions for $p(d|\psi)$ in \eqref{eq:max_p_d_psi} and correspondingly may lead to more efficient methods for source power estimation from the samples in $\sss$.}\par

\added[]{In the presence of more than one acoustic source at the intended frequency, MFP and CNN-based methods may fail in localization. In such scenarios, the JSEA method might show spurious peaks even in an environment that closely matches the training environment, i.e., JSEA reveals a mismatch in the situation (multi-source situation during inference versus the single-source situation during the training). Nevertheless, JSEA is not able to rectify such errors, as it requires samples from the \emph{same} source at different ranges to implicitly estimate the source power.}

\added[]{The simulations in this paper verify the efficacy of JSEA for narrowband sources. For broadband source localization, there are several approaches to exploit the information contained in different frequency bands. One approach is using the SCMs obtained for multiple frequencies as inputs of size $2N_F \times L \times L$ to the CNN model, where $N_F$ is the number of frequencies. In this scenario, JSEA can be used in the same manner it is used the narrowband case. However, another approach is to train a different sub-model for each frequency and then aggregate their estimates to obtain the source range. In this scenario, we can either apply JSEA to each sub-model first and then aggregate the results, or first aggregate the results (e.g., by averaging the output PMFs from all sub-models) and then use JSEA as in the narrowband scenario. The efficacy of this approach remains to be explored in future work.}

\section{Conclusion} \label{sec:conclusion}
\replaced[]{In this paper, we tackled test-time adaptation of a deep-learning based localization algorithm to environmental mismatch between training and test data. We considered different forms of mismatch including that in water depth, bathymetric shape and sediment type}{}. We showed that an appropriate model uncertainty such as MUMI has the potential to reveal the mismatch between the training and test environments. \replaced[]{S}{However, s}ince regression models equipped with MUMI are computationally complex, we resort to a classification paradigm, \replaced[]{using}{where we used} a special label softening to tailor the method to localization. By using classification for localization, the model outputs reveal the peakwise uncertainty, which we then use to enhance the localization based on the source power.\par
  
We showed that while SHOT can slightly improve the performance by encouraging the network to make its predictions less uncertain on the samples that are similar to the training set, JSEA can further improve the UWA localization performance on the samples that suffer more from the environmental mismatch. To be successful, the proposed JSEA method requires a pre-trained model that can  correctly classify part of the test data and do so with certainty.

This paper shows that by suitable domain adaptation, one can enhance localization robustness in data-driven methods. With more prior knowledge about domain shifts between the training and testing environments and source location distribution, one may be able to improve the performance by using more effective methods for selecting $\sss$ even without access to training data. However, the performance improvement will be limited in the absence of such knowledge.  Future potential directions on this front include\replaced[]{ extending the method to broadband sources or multiple sources,}{} investigating other TTA methods currently used for other machine learning problems and developing other methods\added[]{ for} uncertainty quantification in the stated UWA problem.\par

\appendices
\added[]{\protect \section{Complexity Analysis}\label{app:complexity}}

\begin{figure}[htb]
    \centering
    \includegraphics[width=\linewidth]{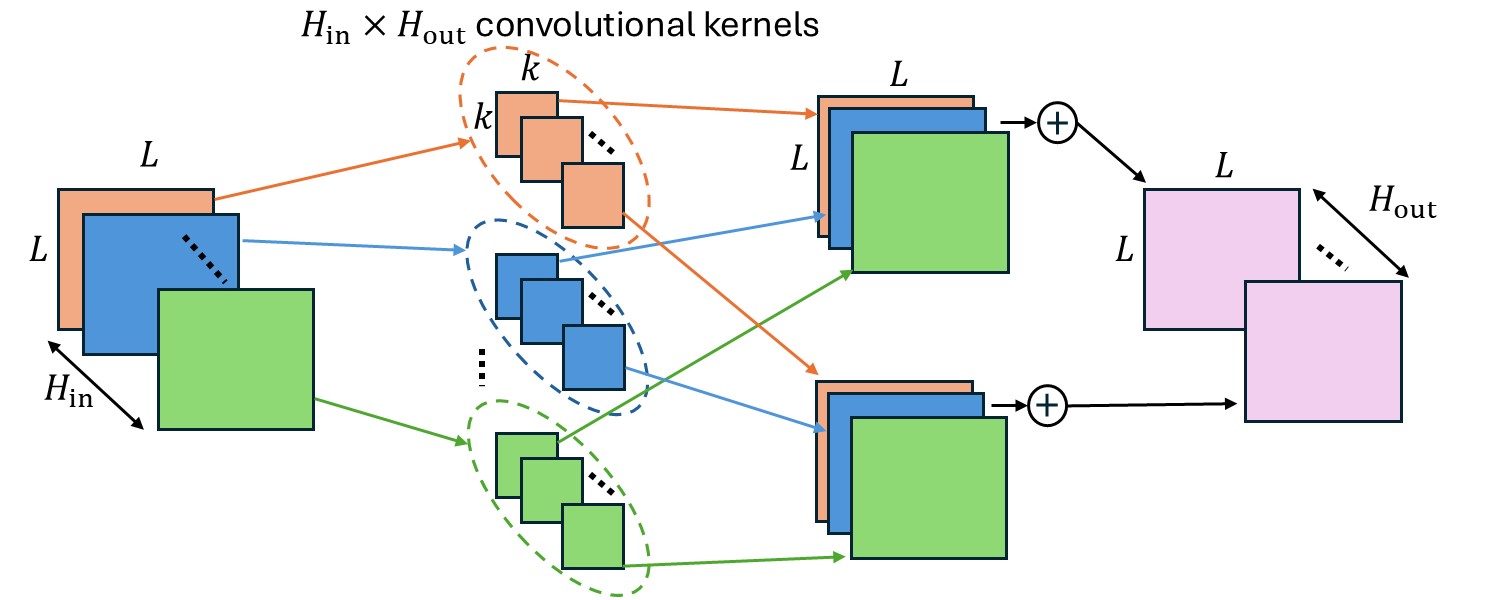}
    \caption{\added[]{A typical convolutional layer with $H_{\text{in}}$ input channels and $H_{\text{out}}$ output channels.}}
    \label{fig:conv-layer}
\end{figure}

\added[]{
Fig. \ref{fig:conv-layer} shows the size of tensors involved in a typical convolutional layer, where there are $H_{\text{in}}$ input channels, there are $H_{\text{out}}$ output channels, and the input and output dimensions are both $L \times L$ due to to padding. According to this figure, the number of multiplications in a convolutional layer is $H_{\text{in}} H_{\text{out}} L^2 k^2$ and the number of additions is $H_{\text{in}} H_{\text{out}} L^2 k^2$, resulting in a computational complexity of $2H_{\text{in}} H_{\text{out}} L^2 k^2$. This yields a computational complexity of $O(L^2)$ and a memory complexity of $H_{\text{in}} H_{\text{out}} L^2$. Moreover, according to Fig. \ref{fig:conv-layer}, each convolutional layer require $H_{\text{in}} H_{\text{out}} k^2$ floating point numbers to be saved, while each linear layer requires a matrix of size input-size $\times$ output-size to be saved. Also, each layer requires an activation memory to hold the input and output tensors. Table \ref{tab:cnn_complexity} summarizes the computational and memory complexities of each layer during a forward pass. In addition, updating the network weights using samples from the test environment involves backpropagation. Each backpropagation iteration requires the same amount of memory that a forward pass needs, to save the gradients with respect to inputs and weights. However, the computational complexity of a backpropagation pass is twice that of a forward pass, as backpropagation involves computing gradients with respect to not only the weights, but also inputs.}

\begin{table*}[h]
\centering
\caption{\added[]{CNN Architecture Complexity Analysis}}
\begin{tabular}{|c|c|c|c|c|c|c|c|}
\hline
Layer & Type & $H_{in}$ & $H_{out}$ & Input Size & Output Size & \makecell{Computational\\Complexity} & \makecell{Memory\\Complexity} \\
\hline
1 & conv (k=3) & 2 & 6 & $L \times L$ & $L \times L$ & $216L^2$ & $2L^2+6L^2 + 108$ \\
\hline
2 & conv (k=5) & 6 & 38 & $L \times L$ & $L \times L$ & $11{,}400L^2$ & $38L^2+5{,}700$ \\
\hline
3 & conv (k=5) & 38 & 40 & $L \times L$ & $L \times L$ & $76{,}000L^2$ & $40L^2+38{,}000$ \\
\hline
4 & linear & $1$ & $1$ & $40L^2$ & $N_{\phi}$ & $80N_{\phi}L^2$ & $40L^2N_{\phi}$ \\
\hline
5 & linear & $1$ & $1$ & $N_{\phi}$ & $M$ & $2N_{\phi}M$ & $MN_{\phi}$ \\
\hline
\end{tabular}
\label{tab:cnn_complexity}
\end{table*}

\added[]{The JSEA methods refines the outputs of the CNN to achieve the final estimates for the source ranges. While the aforementioned complexity analysis applies to any general CNN architecture, the JSEA refinement step requires saving the estimates and energy levels of the received signals for a subset of the test samples (certain samples). Therefore, this refinement step increases the memory complexity by $O(N_{\text{test}})$ floating point numbers. Furthermore, peak finding algorithm may be performed with $O(MW)$ complexity, where $W$ is the size of the sliding window over which we find the local maximum and $M$ is the number of classes. Also, JSEA involves computing the received signal power, which adds to the complexity by $O(M)$ per sample. Finally, for an uncertain sample with $N_P$ significant peaks, we need to construct $\sss_\delta(d) \subseteq \sss$ for each peak and compare the average received power over that set to that of the uncertain sample, which can be done with a complexity of $O(N_P|\sss|)$. Note that while JSEA refinement provides final range estimates, fine-tuning the model (model adaptation) using the pseudo-labeled certain samples is a necessary step in SHOT and involves backpropagation for $N_{\text{itr}}$ iterations. Since each backpropagation iteration is twice as complex as a forward CNN pass, this model adaptation step has a complexity of $2 N_{\text{itr}}$ times that of the forward pass, as mentioned in Table \ref{tab:complexity_comparison}.}

\added[]{Matched field processing method in \eqref{eq:MFP} involves two complex matrix multiplications.
Since $\mathbf{\Tilde{r}}(f,d)$ is of size $L \times 1$, $d \in \{d_1, ..., d_{N_{\text{tr}}}\}$, and $\Tilde{\mathbf{C}}$ is of size $L \times L$, this yields a memory complexity of $2N_{\text{tr}}L$ floating point numbers and a computational complexity of $N_{\text{tr}}(8L^2+8L)$. Table \ref{tab:complexity_comparison} summarizes the complexities of different modules.}

\begin{table*}[h]
\centering
\caption{\added[]{Memory and Computational Complexity Comparison}}
{\small
\begin{tabular}{|c|c|c|}
\hline
Model & \makecell{Computational\\Complexity} & \makecell{Memory\\Complexity} \\
\hline
MFP & $N_{\text{tr}}(8L^2+8L)$ & $2N_{\text{tr}}L$ \\
\hline
\makecell{CNN-c\\forward pass} & $(87,616 + 80 N_{\phi})L^2 + 2MN_{\phi}$ & $(86+40 N_{\phi})L^2 + MN_{\phi} + 43,808$ \\
\hline
\makecell{JSEA-c\\refinement} & $O(M+MW+N_P|\sss|)$ & $N_{\text{test}}(L+1)$ \\
\hline
\makecell{JSEA-c or SHOT\\model adaptation} & $2N_{\text{itr}}((87,616 + 80 N_{\phi})L^2 + 2N_cN_{\phi})$ & $N_{\text{test}}L + (86+40 N_{\phi})L^2 + MN_{\phi} + 43,808$ \\
\hline
\end{tabular}
}
\label{tab:complexity_comparison}
\end{table*}


\bibliographystyle{IEEEtran}
\bibliography{myrefs}


\begin{IEEEbiography}[{\includegraphics[width=1in,height=1.25in,clip,keepaspectratio]{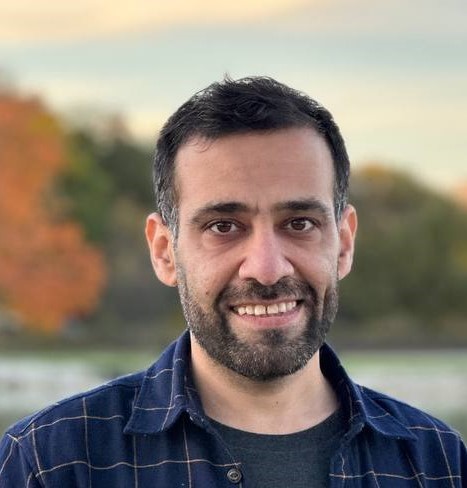}}]{Dariush Kari} is currently a Postdoctoral Research Associate with the Interdisciplinary Health Sciences Institute at the University of Illinois Urbana-Champaign. He received his Ph.D. in Electrical and Computer Engineering from the University of Illinois Urbana-Champaign, USA, 2024, and his M.Sc. in Electrical and Electronics Engineering from Bilkent University, Ankara, Turkey, 2017, and his B.Sc. in Electrical Engineering and in Computer Science, both from Amirkabir University of Technology, Tehran, Iran, 2014. His current research interests include statistical signal processing, machine learning, computational acoustic sensing and imaging, photoacoustic imaging, and underwater acoustics.
\end{IEEEbiography}

\begin{IEEEbiography}[{\includegraphics[width=1in,height=1.25in,clip,keepaspectratio]{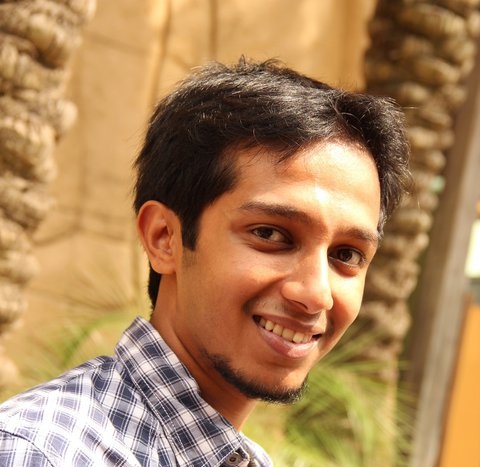}}]{Hari Vishnu} is a Senior Research Fellow at the Acoustic Research Laboratory, National University of Singapore. His interests include machine learning for underwater applications, bio-acoustics and processing in impulsive noise. These are used in a wide range of underwater applications ranging from biodiversity or defense-related scenarios in shallow tropical waters infested with snapping shrimp noise, to polar ice sheets where glacier melt noise dominates the soundscape.

From 2019, Hari is focusing on the acoustics of melting ice, machine-learning based marine-mammal quantification, and distributed acoustic sensing. He obtained his Ph.D from Nanyang Technological University, Singapore, in Computer Engineering on underwater signal processing including robust detection and localization. 

He is the Chief Editor on the IEEE OES Science outreach magazine Earthzine and serves on the IEEE Oceanic Engineering society Executive committee as Deputy Secretary. In 2019, he was awarded the IEEE OES YP-BOOST award which aims to encourage young professionals to participate in the society leadership. 
\end{IEEEbiography}

\begin{IEEEbiography}[{\includegraphics[width=1in,height=1.25in,clip,keepaspectratio]{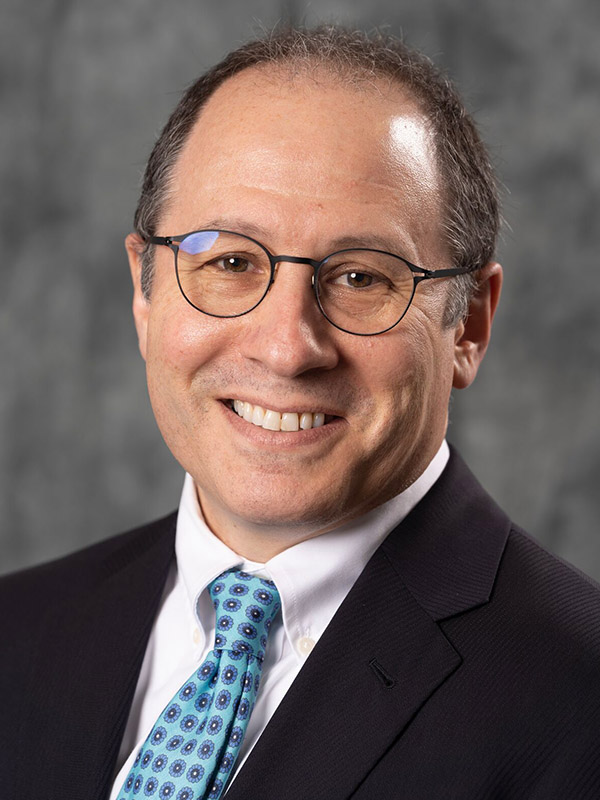}}]{Andrew C. Singer} received the S.B., S.M., and Ph.D. degrees in electrical engineering and computer science from the Massachusetts Institute of Technology (MIT), Cambridge, MA, USA. Since 2023, he is Dean of the College of Engineering and Applied Sciences at Stony Brook University and Professor of Electrical and Computer Engineering. From 1998 to 2023
he was on the Faculty of the Department of Electrical and Computer Engineering at the University of Illinois Urbana-Champaign, where he currently is Fox Family Professor Emeritus. During the academic year 1996, he was a Postdoctoral Research Affiliate with the Research Laboratory of Electronics, MIT. From 1996 to 1998, he was a Research Scientist with Sanders (a Lockheed Martin Company), Manchester, NH, USA.
In 2000, he co-founded Intersymbol Communications, Inc., and in 2014, he co-founded OceanComm, Inc.
In 2009, he was elected Fellow of the IEEE “for contributions to signal processing techniques for digital communication,” and in 2014, he was named a Distinguished Lecturer of the IEEE Signal Processing Society.
\end{IEEEbiography}

\end{document}